\documentclass[preprint2]{aastex}
\usepackage{graphicx}% Include figure files
\usepackage{dcolumn}% Align table columns on decimal point
\usepackage{amsmath}
\usepackage{bm}% bold math
\usepackage{float}
\renewcommand{\vec}[1]{{\mbox{\boldmath$#1$}}} % Bold
\newcommand{\mat}[1]{\ensuremath{\mathbf #1}}   % Bold upright for tensors
\newcommand{\inv}[1]{\ensuremath{{\mathbf #1}^{-1}}}   % Bold upright for tensors

\newcommand{\dif}    {\ensuremath{\,\mathrm{d}}} % Math differential qty

\newcommand{\be}{\begin{equation}}
\newcommand{\ee}{\end{equation}}
\newcommand{\ba}{\begin{align}}
\newcommand{\ea}{\end{align}}

\shorttitle{A Multi-Pulse Fitting Technique}

\begin{document}

\title{
  Microcalorimeter Spectroscopy at High Pulse Rates:\\ a Multi-Pulse 
  Fitting Technique
}

\author{J.W. Fowler\altaffilmark{1}, B.K. Alpert\altaffilmark{1}, W.B. Doriese\altaffilmark{1}, D.A. Fischer\altaffilmark{2},
C. Jaye\altaffilmark{2}, Y.I. Joe\altaffilmark{1}, \\ G.C. O'Neil\altaffilmark{1}, D.S. Swetz\altaffilmark{1}, J.N. Ullom\altaffilmark{1}}
\altaffiltext{1}{National Institute of Standards and Technology, 325 Broadway MS\,686.02, Boulder, Colorado 80305, USA}
\altaffiltext{2}{National Institute of Standards and Technology, Brookhaven National Lab, Brookhaven, New York, USA}

%\email[Electronic mail: ]{joe.fowler@nist.gov}

%\author{D.A. Fischer, C. Jaye}
%\affil{National Institute of Standards and Technology, Brookhaven National Lab, Brookhaven, New York, USA}
%
%\author{J. Uhlig}
%\affil{Department of something, Kemicentrum, Lund University, Lund, Sweden}

%\nistauthor{G.C. Hilton}
%
%\nistauthor{K.D. Irwin}
%
%\nistauthor{C.D. Reintsema}
%
%\nistauthor{D.R. Schmidt}

%\nistauthor{L.R. Vale}

\date{\today}

\begin{abstract}

% JWF: Abstract should be <~100 words

  Transition edge sensor microcalorimeters can measure x-ray and gamma-ray energies with very high energy resolution and high photon-collection efficiency.  For this technology to reach its full potential in future x-ray observatories, each sensor must be able to measure hundreds or even thousands of photon  energies per second.  Current ``optimal filtering'' approaches to achieve the  best possible energy resolution work only for photons well isolated in time, a  requirement in direct conflict with the need for high-rate measurements.  We describe a new analysis procedure to allow fitting for the pulse height of all photons even in the  presence of heavy pulse pile-up.  In the limit of isolated pulses, the technique reduces to the standard optimal filtering with long records. We employ reasonable approximations to the noise covariance function in order to render multi-pulse fitting computationally viable even for very long data records.  The technique is employed to analyze x-ray emission spectra at 600\,eV and 6\,keV at rates up to 250 counts per second in microcalorimeters having exponential signal decay times of approximately 1.2\,ms.
  \vspace{10mm}
\end{abstract}

\maketitle

%=============================================================================
\section{Motivation and goals}
\label{sec:introduction}
%=============================================================================

The Transition Edge Sensor (TES) is an excellent calorimetric spectrometer with applications in 
physical chemistry, materials analysis, gamma-ray spectrometry for nuclear forensics, and x-ray 
astronomy, among many other uses.  A variety of satellites and sounding rockets are under development or 
have been proposed as platforms for TESs or other cryogenic microcalorimeters (to which our results 
also apply).  Currently, an array of nearly 4000 TESs is planned as the focal plane for the X-ray Integral 
Field Unit (X-IFU)  on the recently selected European x-ray satellite \emph{Athena} \citep{Ravera:2014hv}.  

Absorption of a single x-ray photon produces a current pulse lasting for a fraction of a millisecond to a 
few milliseconds under typical TES operating conditions. The current 
state-of-the-art ``optimal filtering" techniques for pulse height estimation work well only when pulses are 
isolated from all earlier and later pulses over an isolation period several times longer than this. Such 
techniques can deal with pulse pile up only by discarding non-isolated pulses. Hence, optimal filtering is 
a good approach in the limit of low photon rates. When the product of the exponential decay 
time of the sensors and the photon count rate approaches and exceeds 0.05 to 0.10, however, 
optimal filtering forces increasingly difficult compromises between photon throughput and energy resolution.  
Relieving this tension through faster time constants is difficult; faster pulses generally require alterations to the 
sensor designs and---more critically---additional readout bandwidth.  
To preserve resolution for sensors with a decay time of 150 microseconds (as the
 \emph{Athena} plan specifies), pileup rejection becomes increasingly wasteful of photons at a pulse rate 
per sensor of 100 counts per second or more.

Instruments such as the \emph{Athena} X-IFU have a wide range of science targets~\citep{AthenaAssessment:2011}. 
Some extended targets such as galaxy clusters and emission from filaments of the Warm Hot 
Interstellar Medium require high-resolution spectroscopy across the full field of view of the X-IFU.  
These targets are particularly well matched to TES sensors.  However, the same 
instrument must also perform spectroscopy on compact objects whose spatial extent at the focal plane 
is set by the point-spread-function of the optics, objects such as neutron star binaries and stellar mass black holes.  
In these cases, the full x-ray flux of the source will be 
incident on one or just a few TESs and the count rate capabilities of the individual sensors will strongly 
affect measurement quality.  The X-IFU specifications require at least 80\,\% of photons to be measured 
with the highest resolution for 1 milli-Crab source intensities.  The count rate from a milli-Crab source 
will depend on parameters  that are still under discussion,  such as mirror area, but this requirement 
corresponds approximately to achieving output count rate of 40 counts per second (cps) per TES at an 
input count rate of 50 cps.  For sources that exceed 1 milli-Crab, the fraction of lost photons will be even 
higher than  20\,\%.   One viewing strategy for bright compact objects is to insert a diffuser that spreads 
the x-rays across a larger area of the focal plane.  Alternatively, improved signal-processing strategies may increase 
photon throughput.  We describe here a pulse-analysis technique that can dramatically advance the 
goal of preserving spectral resolution at high count rates relative to existing methods. 

To perform any given observation as quickly as possible and make efficient use of the highest photon 
fluxes, we must be able to estimate pulse heights in the presence of some modest amount of pileup 
with resolution comparable to that achieved on isolated pulses. Previous x-ray telescopes such as the 
ASTRO-E \citep{Boyce:1999ey} and ASTRO-H missions \citep{Takahashi:2012dh} address the problem 
by ``event grading.''  Pulses not isolated enough to use the best optimal filter use a  shorter ``mid-grade'' 
filter; pulses not suited even for the shorter filter are simply averaged over several successive 
samples near the pulse peak, for ``low-grade'' filtering. Even this method, though, is hampered by 
having minimum pulse-isolation requirements.

We present here a different approach based on linear superposition of pulses. 
Its primary goal is to break the tight link between photon 
throughput and energy resolution that the isolated-photon requirement imposes. This requirement 
arises in the traditional analysis because optimal filtering is designed to consider only one pulse at a 
time; our method relaxes this rule. Our secondary goal is to create a method simple enough and fast 
enough to implement in firmware or hardware, so that it can be performed on a spacecraft. It should 
handle pulse pileup in a way that works at low pulse rates, too, and not rely on different analysis 
modes for high and low rates. It is also important that the method not require substantially more data to 
be sent back from space than a traditional method does, given the limited bandwidth for communication from space; 
the method should yield a pulse height and arrival time, 
and not more than a few additional quantities for each x-ray pulse.

We first describe the standard technique of optimal filtering and show how it generalizes naturally to the 
estimation of multiple pulse heights simultaneously in a long record of sensor data, by a  
form of weighted least-squares fitting (Section~\ref{sec:pulseheights}).  
A straightforward application of this idea 
to long data records would be computationally quite expensive, the cost scaling quadratically with the 
samples per data segment. This high computational cost was noted by an earlier group that applied this 
idea to astronomical microcalorimeter data from a brief suborbital flight 
\citep{Crowder:2012kk}. One possible compromise---to make an implicit 
assumption of white noise---has 
been used to create FPGA-based hardware performing simultaneous pulse-height fits for very fast 
silicon drift diodes \citep{Scoullar:2011dw}. We present a method to reduce the computational complexity of 
the fitting procedure even in the presence of non-white noise by approximating the noise autocorrelation 
function as the result of a low-order autoregressive moving average (ARMA) process (Section~
\ref{sec:fast_noise}). This powerful approximation has not previously been used with x-ray 
microcalorimeter data.

The technique is then applied to two terrestrial TES measurements.  The data 
were recorded by microcalorimeters designed for x-rays up to 10\,keV.  The first data set (Section~
\ref{sec:nsls_data}) employs an array of forty sensors at energies below 1\,keV. Because the x-ray energy 
is far below the saturation energy, the energy-resolving power is modest but the detectors are highly 
linear. The x-ray source is the beamline U7a at the Brookhaven National Laboratory's National Synchrotron 
Light Source (NSLS).  This example application is chosen to have both high photon rates and minimal 
nonlinearity.

\begin{figure}[ht!]
\includegraphics[width=3.125in]{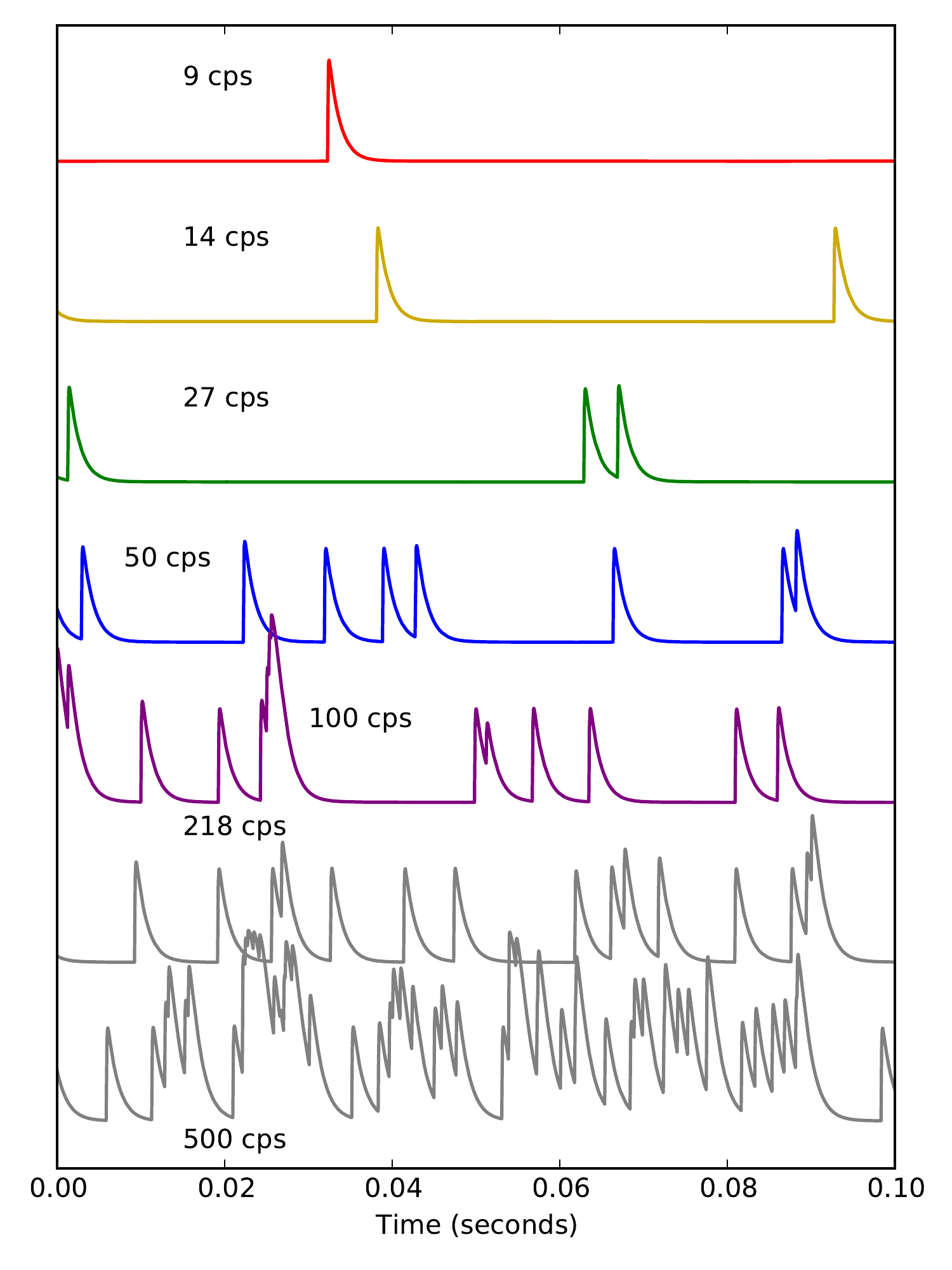}
\caption{\label{fig:example_highrate}
  Example microcalorimeter signals, taken from one TES sensor observing 6\,keV fluorescence x-rays, 
as described in Section~\ref{sec:mn_data}.  Each curve shows 100\,ms of raw data corresponding to a 
different x-ray source intensity, producing the indicated number of counts per second (cps) in the sensor.  The 
deleterious effects of sensor nonlinearity prevent use of the 218 and 500 cps data in this work, but the data up to 
100 cps are analyzed here.
   }
\end{figure}

The other data set (Section~\ref{sec:mn_data}) consists of the 5.9\,keV and 6.4\,keV fluorescence emission of a 
manganese target illuminated by an x-ray tube source.  The emission is measured by a single detector over a range 
of photon rates from 9 to 100 photons per second. Figure~\ref{fig:example_highrate} shows 100\,ms of 
observation from each of the five data rates, giving a sense of the problem being addressed.
The TES calorimeters are less linear at these energies than they 
are below 1\,keV, while detector linearity is an assumption intrinsic to our method. Still, the multi-pulse fitting method 
shows impressive performance even at high rates with only the simplest of corrections for nonlinearity.  High-rate 
astronomical spectroscopy will present problems due to sensor nonlinearity that are beyond the scope of this work, 
though in Section~\ref{sec:conclusions} we offer some suggestions on ways to move beyond the technique featured here.

%=============================================================================
\section{Pulse height estimation from optimal filtering to multi-pulse fitting}
\label{sec:pulseheights}
%=============================================================================

%improvement is minimal beyond a certain length).  Longer filters lead to more
%severe requirements on the isolation of pulses in time, however.  In short, we must
%sacrifice more pulses to pile-up cuts to improve the energy resolution.  This trade
%off becomes intolerable at higher pulse rates.
%

The data used for estimating microcalorimeter pulse amplitudes consist of the regularly sampled and digitized values of the time-varying bias current through the sensor. The sample rate is typically $10^5$ to $10^6$ samples per second per TES. In the \emph{Athena} X-IFU---an array of 4000 sensors---1~GB of raw data will be generated in each second of operation, or nearly 100~TB per day.  While this high data rate demands fast analysis procedures, the extremely high resolving power of the microcalorimeters also means that one must take great care not to lose information or to introduce new systematic errors. Balancing the competing requirements of \emph{computational} and \emph{statistical} efficiency is among the central challenges in analyzing TES data.

We distinguish here between estimating the \emph{amplitude} of the TES current pulse and the \emph{energy} of the absorbed photon, not only because the scale factor between the two is not known a priori, but also because the relationship is generally not a linear one. Understanding this nonlinear mapping is an important concern in analyzing TES data, but it is beyond the scope of the current paper. We will assume for now that precise and accurate estimates of pulse amplitudes are a necessary and sufficient condition for making similarly good estimates of pulse energy and focus on the problem of amplitude estimation.

%---------------------------------------------------------------------------------------------------------------------
\subsection{Optimal filtering of isolated pulses}
\label{sec:opt_filtering}

%\emph{Briefest possible derivation of standard optimal filtering, then constrained optimal filtering. Ugh: have to find the right citations.}

The standard technique for estimating microcalorimeter pulse amplitudes begins with the identification of pulses and their arrival times from a long stream of regular TES current samples (``triggering''), and the subsequent selection of a predetermined number of samples both before and after the pulse onset (generating a ``pulse record'' surrounding the trigger time). The amplitude estimate for each pulse is then a certain weighted sum of the values in the pulse record. The best set of weights is the so-called \emph{optimal filter} 
\citep{Szymkowiak:1993ck,Boyce:1999ey,Lindeman:2000ta}. This filter can be derived in various ways---as a minimum-variance unbiased estimator, or as a maximum-likelihood estimator---starting from these assumptions:
\begin{enumerate}
\item Pulses always arrive long after any energy deposited by earlier x-rays has fully dissipated. The sensor has returned to its steady-state electrical and thermal conditions.
\item Pulses are all identical in shape, regardless of energy. Only a single scale factor---the pulse amplitude---varies from one pulse to another.
\item The pulse shape is known ahead of time, perhaps by averaging together many pulses from an initial training portion of the data.
\item Noise is additive and follows a multivariate Gaussian distribution.
\item The noise is stationary (in particular, it does not vary as the current in the sensor evolves), and its autocorrelation function (or equivalently, its power spectral density) is  known.
\end{enumerate}
Not one of these assumptions is strictly correct, yet the procedure yields splendid results in a wide variety of conditions, at least for time-isolated pulses. In this paper, we consider violations of the first condition: cases where some measurable amount of residual energy remains in a sensor when a new pulse arrives. This problem is known as \emph{pulse pile-up}. Our approach to the problem is to assume strict sensor linearity, so that the current in the sensor is  simply the sum of the decay to equilibrium that would have been seen in the absence of the new pulse, plus the new pulse as it would have been measured in isolation. This assumption, of course, is also incorrect;  we will explore how far we can go with it, nevertheless.

Returning to the standard analysis, let \mat{R} be the covariance matrix of the noise. That is, its elements are expectation values (here, $E[\cdot]$) of two-point products for a pulse-free data sequence $\{d_i\}$,
\[
R_{ij} = \mathrm{E}[d_i d_j] - \mathrm{E}[d_i]\ \mathrm{E}[d_j].
\]
Because the noise is assumed to be stationary, the symmetric matrix \mat{R} is Toeplitz, i.e., $R_{ij}=r_{|i-j|}$ for some sequence $r_t$, the noise autocovariance function. A symmetric Toeplitz matrix is fully specified by its first row or column. 

Let the assumed pulse shape, called the \emph{pulse model}, be the vector\footnote{We treat all vectors as column vectors.} \vec{s} and the measured data be the vector \vec{d}. We choose the length $N$ of both vectors in advance (typically by fixing the record length at data-acquisition and triggering time).  In the simplest case, there is nothing expected in the data but the pulse of unknown amplitude $p$ plus zero-mean noise, i.e.
\[
\mathrm{E}[\vec{d}] = p\vec{s},
\]
and it can be shown that the best (minimum-variance, unbiased) estimate of the pulse height is
\[
\hat{p} = (\vec{s}^T\inv{R}\vec{s})^{-1} \vec{s}^T \inv{R} \vec{d} \equiv \vec{\hat{f}}_1^T \vec{d}.
\]
Here we have introduced $\vec{\hat{f}}_1$, the optimal filter for estimation of a single amplitude. 
It is a weighting vector, which lets us estimate $\hat{p}$ by taking an inner product with the data vector.

Most cases are not so simple, and one must account for additional terms in the data other than noise and the pulse itself. One specific term that must be included in nearly every analysis is an additive constant, called the \emph{baseline}. This constant arises because systems for measuring the tiny currents through microcalorimeters generally have unknown (and slowly varying) DC offsets.  Removing the DC term specifically is straightforward \citep{Doriese:2009AIPC}, but we prefer a more general treatment. We have previously introduced a framework for \emph{constrained optimal filtering} \citep{Alpert:2013RScI}, in which the filter is the minimum-variance filter chosen not from the set of \emph{all} possible weights, but only from those that are strictly insensitive to one \emph{or more} additive ``nuisance terms''. In this picture, we replace the vector \vec{s} with a model matrix \mat{M} whose first column is \vec{s} (the pulse shape); other columns contain all anticipated nuisance terms. One such column might be a column of ones, if DC-insensitivity is required; another might be a decaying exponential chosen to render the filter insensitive to the tail of an earlier pulse (provided its decay time is known). In this approach, the best estimate of the pulse height becomes
\be \label{eq:std_filter}
\hat{p} = \vec{e}^T_1(\mat{M}^T\inv{R}\mat{M})^{-1} \mat{M}^T \inv{R} \vec{d} \equiv \vec{\hat{f}}^T \vec{d}.
\ee
Here we use the unit vector $\vec{e}_1\equiv[1,0,0,\dots]^T$ to select only the term that corresponds to the amplitude of the pulse shape component \vec{s} and to discard all other terms.

It can be shown that each new component introduced into the model increases the variance of the estimator $\hat{p}$, but in practice the signal-to-noise price paid is often quite small, particularly when the records are long or the nuisance terms are very different from the pulse shape \vec{s}.

These results use the noise covariance matrix $\mat{R}$ rather than the noise power spectrum. Use of the Fourier space representation of noise is both possible and convenient, but it introduces an additional wrong assumption of periodicity; in fact, neither the finite-length pulse nor the noise in a record are periodic. The signal-to-noise cost for filtering in the Fourier domain may be small in some cases \citep{Alpert:2012JLTP}, but we prefer to avoid paying unnecessary costs, however small.

%---------------------------------------------------------------------------------------------------------------------
\subsection{Filtering as fitting for pulse amplitude}
\label{sec:fitting_1pulse}

A different and productive view of the optimal filter (Equation~\ref{eq:std_filter}) is also possible.  Suppose, as before, that we are given the noise model (the matrix \mat{R}), the pulse shape (\vec{s}), and a complete set of components in a model of the pulse records (the columns of matrix \mat{M}). If we also knew the components' true amplitudes \vec{p}, we could compute the likelihood, the probability under the linear model that the data would be \vec{d}. Ignoring normalization factors, the likelihood is
\[
\mathcal {L} \propto \exp\left[ -(\vec{d}-\mat{M}\vec{p})^T \inv{R} (\vec{d}-\mat{M}\vec{p}) \right ].
\]
For a fixed measurement \vec{d} and model, the logarithm of $\cal L$ is a quadratic function of the unknown parameters \vec{p}. If we maximize the likelihood with respect to the vector \vec{p} by setting 
\newcommand{\diff}{\ensuremath{\mathrm{d}}}
\[
\diff{(\log\mathcal{ L})}/\diff{\vec{p}} |_{\vec{p}=\vec{\hat{p}}} =0,
\]
then we find a closed-form expression for the maximum-likelihood estimates of the parameters:
\be \label{eq:max-like}
\vec{\hat{p}} = (\mat{M}^T\inv{R}\mat{M})^{-1} \mat{M}^T \inv{R} \vec{d}, % \equiv \vec{\hat{f}}^T \vec{d},
\ee
of which Equation~\ref{eq:std_filter} is one component. Note that the estimate is unbiased:
\[
\mathrm{E}[\vec{\hat{p}}] = \vec{p}
\]
because $\mathrm{E}[\vec{d}] = \mat{M}\vec{p}$.
 This establishes that the minimum-variance estimator for the pulse height quoted in the previous section is in fact also the maximum-likelihood estimate of the pulse height, provided that one fits simultaneously for the amplitudes of all components of the model (including the baseline level, unwanted pulse tails, and so forth).

The covariance of our parameter estimates can also be computed:
\[
\mat{C}_{pp} = \mathrm{E}[ (\vec{\hat{p}}-\vec{p}) (\vec{\hat{p}}-\vec{p})^T  ] =
\mathrm{E}[\vec{\hat{p}} \vec{\hat{p}}^T] - \vec{p} \vec{p}^T.
\] 
The notation is simplified if we designate the system \emph{design matrix} as
\be \label{eq:design}
\mat{A} \equiv \mat{M}^T\inv{R}\mat{M}.
\ee
Then
\begin{align}
\mat{C}_{pp} &= \inv{A}\mat{M}^T\inv{R}\mathrm{E}[\vec{d}\vec{d}^T] \inv{R}\mat{M}\inv{A} - \vec{p} \vec{p}^T \nonumber \\
&=  \inv{A}  (\mat{M}^T\inv{R}\mat{R} \inv{R}\mat{M})  \inv{A} \label{eq:param_covar_long}\\
&=  \inv{A} \mat{A}\inv{A} \nonumber \\
&= \inv{A}. \label{eq:param_covar}
\end{align}
This estimate of the parameter covariance  will be useful in what follows.\footnote{Equation~\ref{eq:param_covar_long} can also be useful in the event that we wish to perform filtering with an \emph{approximate} noise covariance matrix \mat{R}, but we believe the \emph{true} covariance to be \mat{T}. See Section~\ref{sec:effect_noise_approx}.}

%---------------------------------------------------------------------------------------------------------------------
\subsection{First-order treatment of the variation in x-ray pulse arrival times}
\label{sec:arrival_time}

%\emph{Simply mention that one can add terms to the model in order to accommodate sub-sample variations in arrival time, if we model those variations as simple shifts and if we work only to first order. This will be needed to get our best possible result in Section~\ref{sec:mn_data}.  I'm not sure, but this section might want to go before Section~\ref{sec:fitting_multiple}.}

One source of systematic error in pulse-height analysis is the undesirable dependence of the estimated pulse heights on the exact arrival time of an x-ray photon. Photons arriving at different times with respect to the TES current-sampling clock will appear to have slightly different pulse shapes, and this effect is often large enough to degrade the energy resolution if left uncorrected.

A full cancellation of this arrival-time effect presents a major unsolved challenge to developers of pulse-processing techniques. Nevertheless, we find that the effect can be suppressed substantially by a first-order treatment. We assume that the pulse model \vec{s} is the result of regularly sampling an unknown, smooth ``pulse shape'' function $f(t)$ with a sampling period of $\Delta$:
\[
s_i = Af(i\Delta-t_\mathrm{a}), \ i={...,-1,0,1,2,...}
\]
where $A$ is the pulse amplitude, and $t_\mathrm{a}$ is the unknown exact arrival time of the photon. We can treat $t_\mathrm{a}$ as a parameter of the pulse model; unfortunately, the model is nonlinear in this parameter, invalidating the analysis of the previous section. If the function $f(t)$ is expanded in a Taylor series about $t_\mathrm{a}=0$ and only the linear-order term is kept, however, then the result is necessarily linear:
\[
s_i \approx Af(i\Delta) - At_\mathrm{a} f^\prime(i\Delta) + {\cal O}(At_\mathrm{a}^2 f^{\prime\prime}).
\]
In this approximation, we can treat $A$ and $(At_\mathrm{a})$ as the two coefficients of a linear model and solve for them simultaneously by using the model pulse and its time derivative as two of the columns of model matrix \mat{M} and solving Equation~\ref{eq:max-like}. For this purpose, the time derivative $f^\prime(t)$ of a pulse is not known exactly, but it can be approximated by the first finite difference of the pulse model \vec{s}.

We find that this first-order treatment of the variation in photon-arrival times is necessary to achieve the highest possible resolution in the test cases described below in Sections~\ref{sec:nsls_data} and \ref{sec:mn_data}.

%---------------------------------------------------------------------------------------------------------------------
\subsection{Fitting for multiple pulse heights simultaneously}
\label{sec:fitting_multiple}

% The figure demonstrating MPF
\begin{figure*}[ht]
\includegraphics[width=\textwidth]{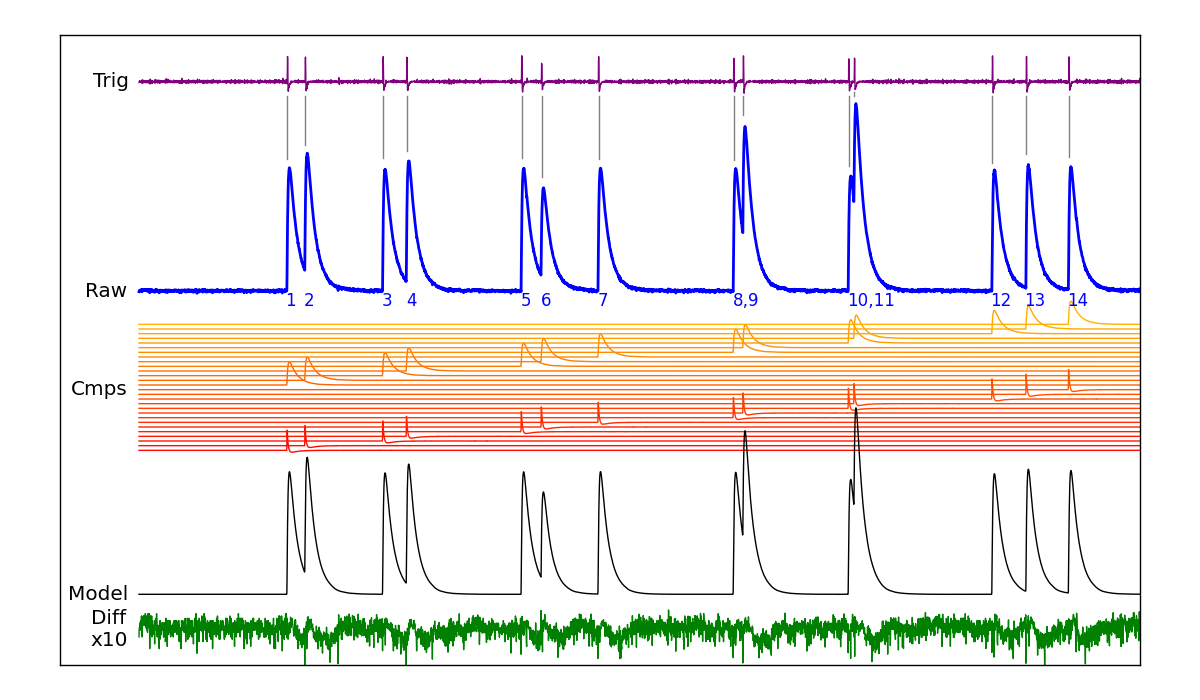}
\caption{\label{fig:mpf_example}
  Multi-pulse fitting requires modeling the raw data (``Raw'', blue) as a
   linear combination of two components per pulse (``Cmps''), identical except for
  different pulse arrival times.  One component is the pulse model $p(t)$, and the other is the first finite time difference of the pulse model (representing $\dif p/\dif t$).  The top curve (``Trig'', purple) represents the raw data convolved with the high-pass kernel used to identify the onset of new pulses. The 14 pulse times are shown by vertical gray lines.
  The best-fit model (``Model'', black)
  and the data minus model residual \emph{times ten} (``Diff x10'', green) are also shown.}
\end{figure*}

Within the assumptions of the linear model, $\vec{\hat{p}}$ is the maximum-likelihood estimator for the amplitudes of all components in the model simultaneously.
To this point, we have supposed that the columns of \mat{M} consist of one pulse shape and multiple nuisance terms such as a constant baseline and possibly the pulse's time derivative. We need not restrict the model to contain only one pulse, though. We can instead analyze a segment of data known to contain two or more pulses. Equation~\ref{eq:max-like} will handle this case equally well;  in such situations, more than one parameter of $\vec{\hat{p}}$ will estimate a pulse height of interest.

This approach, which we call \emph{multi-pulse fitting}, allows much greater flexibility in handling pulses that are not cleanly separated in time. Rather than simply rejecting both members of a closely spaced pair, we can determine the linear combination of two pulses that best fits the measurements.  How close in time two pulses can be without invalidating the estimates is now a decision that can be made at analysis time instead of at data-acquisition time. In any event this approach is much more tolerant of modest pileup.

Figure~\ref{fig:mpf_example} illustrates the procedure.
The example data (labeled ``Raw'' in the figure) depict a 73\,ms segment (5700 samples) from an observation of the fluorescence of NH$_4$NO$_3$ illuminated by a monochromatic 542\,eV x-ray beam.  X-ray photons struck this particular detector at an average rate of 215 counts per second (the array-wide rate was over 8000 cps); 15.6 are expected and 14 are seen during this example.   Assuming that the noise covariance and the expected pulse shape have already been determined during a ``training pass'' through some or all of the data, the first step in analyzing this record is to locate the pulse arrival times.  For this purpose, the data are convolved with a filter having weights $[+1, -.8 -.5, -.2, +.1, +.4]$. This has the effect of finding the difference between each data sample and the extrapolation of the least-squares fit of a line to the five previous samples.  This convolution (``Trig'') is checked for large positive excursions, which define the trigger times.  In this record, 14 photons are found.  The matrix \mat{M} is constructed with 5700 rows (the record length) and 29 columns. One column is the constant value 1 (representing the unknown baseline level); 14 are the standard pulse model, each offset in time by the appropriate trigger time; and 14 are the finite-difference approximation to the pulse model's first time derivative (used as described in the previous section to allow for a first-order correction to the exact pulse arrival time).  For convenience, the standard pulse model is scaled to have exactly unit height, which ensures that the estimate of the amplitude has the same arbitrary units as the raw data.  Given this \mat{M} and an estimate of the noise \mat{R}, Equation~\ref{eq:max-like} can be solved for the estimated parameters, yielding 14 pulse-height estimates, 14 arrival-time corrections, and one constant baseline level. We find that all photons are 525\,eV oxygen fluorescence except for the sixth, which is a 397\,eV nitrogen fluorescence x-ray.  The best fit to the data, given by $\vec{\hat{d}}\equiv \mat{M}\vec{\hat{p}}$, is shown (``Model''), followed by the data-minus-model difference, scaled by a factor of 10.

The parameter-uncertainty estimates (Equation~\ref{eq:param_covar}) predict for most pulses a $1\sigma$ pulse-height uncertainty of 0.9\,eV (or 2.0\,eV full-width at half-maximum, FWHM). When pulses are piled up, these uncertainties increase, but only slightly. Pulses 8 and 9 have 2.1\,eV predicted FWHM, while pulses 10 and 11, although separated by only 0.4\,ms, have 2.4\,eV\@.  The matrix $\mat{C}_{pp}$ also predicts correlations among the estimates of $\vec{\hat{p}}$. The Pearson correlation coefficient between well separated pulses is typically $+0.01$, though closely spaced pulses can have strong anticorrelations.  Correlations of $-0.3$ and $-0.5$ are found between pulses 8 and 9 and between 10 and 11, respectively.  All pulse heights are anti-correlated with the baseline estimate, with $\rho=-0.13$ in all cases.

% DAQ-time choices
It should be apparent that multi-pulse fitting requires a different strategy for selecting segments of data to store for later processing. Instead of implementing a software trigger and recording fixed-length records appropriate for the duration of a single pulse, the data-acquisition software needs to allow for longer records whenever pulses pile up on one another. For the test cases explored here, we have taken this approach to its extreme limit and simply recorded \emph{all} data samples without interruption. This choice consumes the maximum possible storage space but has the advantage of simplicity during data acquisition.

% How to break up data?
When the online system does not separate the data naturally into distinct records,  the offline analysis system must do so. The approach used here is to choose a default record length, typically 10 to 50\,ms. When a pulse occurs too close to the end of such a record, the record is shortened (and the next record lengthened) as needed to ensure that the pulses in any one record have minimal coupling with the next record. At extremely high pulse rates, no gaps of satisfactory size would occur, and a different solution would be required.

%=============================================================================
\section{Through the noise-whitening bottleneck}
\label{sec:fast_noise}
%=============================================================================

Important practical difficulties can arise when  Equation~\ref{eq:max-like} is used for multi-pulse fitting.  
Data sets can contain very many records per second, each of length $N\sim10^4$ or $10^5$.
Noise covariance matrices of size $N\times N$ are costly to invert; for $N\gtrsim 10^5$, it can be impossible even to store \mat{R} or \inv{R}.
Because of the Toeplitz symmetry,  the Levinson algorithm \citep{NumRec3:2007} can be used to solve \mat{R} in $O(N^2)$ time with  only $O(N)$ storage. This is a great improvement over general matrix solutions, which require $N$ times as much of both time and storage, yet Levinson is too slow for general use in applications involving large TES arrays.
So-called ``super fast Toeplitz solvers'' requiring only $O(N\log N)$ operations have been proposed in recent years \citep{Chandrasekaran:2008ti}. Unfortunately, they have very high break-even points, and mid-size problems like the current one are unlikely to reap the benefits larger problems will from the reduced asymptotic complexity.

The model matrix \mat{M}, whose columns are the components of the data model, is different for every segment of data. While two segments might  have the same number of pulses, their pulses will be at different times; for all practical purposes, these are entirely different data models. Since each data segment will have its own \mat{M},  evaluating the maximum-likelihood estimates of the parameters requires computing $\inv{R}\mat{M}$, or one solution of \mat{R} for each model component for each data segment.

Of course, a standard (Levinson) solution of the full noise covariance matrix will be practical for smaller problems. Readers concerned only with such cases may skip over the remainder of Section~\ref{sec:fast_noise}, which discusses approximations to \mat{R} for more demanding situations.

%---------------------------------------------------------------------------------------------------------------------
\subsection{Noise whitening the data and the model}
\label{sec:mahalanobis}

%\emph{Point out that the solution of the noise matrix \mat{R} can be written equivalently as the application of a noise whitening matrix \mat{W} to both data and model components, where any whitener satisfying $\mat{W}\mat{R}\mat{W}^T=\mat{I}$ suffices. Note that the inverse of the lower Cholesky factor of \mat{R} is one such whitener.}

The expression for the best estimate of the model parameters, Equation~\ref{eq:max-like}, twice uses the inverse of the noise-
covariance matrix. The true noise-covariance matrix of any random process is positive definite.\footnote{An empirical estimate of \mat{R} is \emph{not} guaranteed to be positive definite if the covariances $r_t$ are computed from data in the usual, unbiased way.
A different estimator of \mat{R} can be employed 
that is guaranteed to be positive definite, though it is also biased \citep{BrockwellDavis:2009}. In practice, we have never found it 
necessary to use the biased estimator for the noise covariance.}
It follows that \mat{R} is invertible, that its inverse is also positive 
definite, that both admit a Cholesky decomposition, and that both have strictly positive eigenvalues.

It is therefore possible to construct matrix \mat{W}, which is a linear \emph{noise-whitening operation} in the following sense: the covariance of  a whitened data vector ($\vec{w}\equiv\mat{W}\vec{d}$) is
\ba
\mat{C}_{w} &=
\mathrm{E}[\mat{W}\vec{d}\vec{d}^T\mat{W}^T]-\mathrm{E}[\mat{W}\vec{d}]\ \mathrm{E}[\vec{d}^T\mat{W}^T] \nonumber \\
&= \mat{W}\mat{R}\mat{W}^T. \nonumber
\end{align}
If \mat{W} satisfies
\be \label{eq:whitener}
\mat{W}\mat{R}\mat{W}^T = \mat{I},
\ee
or equivalently
\be \label{eq:whitener2}
\inv{R} = \mat{W}^T\mat{W},
\ee
then the whitened data vector ${\vec{w}}$ is linear in the data and has uncorrelated noise of unit variance.

We write the maximum-likelihood fit for the parameters and the design matrix
with tildes indicating whitened vectors or matrices:
\begin{align}
\vec{\hat{p}} &= \inv{A}\tilde{\mat{M}}^T\tilde{\vec{d}} \\
\mat{A} &= \mat{M}^T\mat{W}^T\mat{W}\mat{M}  \stackrel{\mathsf{def}}{=} \tilde{\mat{M}}^T\tilde{\mat{M}}.
\end{align}
The effect of whitening the data and the model components is to reduce the problem from one of non-white noise (requiring correlated weights) to the simpler problem of an unweighted least-squares fit of a linear model. We stress, however, that these two equations are completely equivalent to Equations~\ref{eq:max-like} and \ref{eq:design}.

At least three straightforward constructions of the whitener satisfy Equations~\ref{eq:whitener} and \ref{eq:whitener2}:
\begin{enumerate}
\item Let $\mat{R}=\mat{P}^T\mat{\Sigma}\mat{P}$ be the eigen-decomposition of \mat{R}, where $\inv{P}=\mat{P}^T$ and $\mat{\Sigma}$ is the diagonal matrix of eigenvalues. Then $\mat{W}_e\equiv \mat{\Sigma}^{-1/2}\mat{P}$ is a whitener.
\item Let $\inv{R}=\mat{U}^T\mat{U}$ be the Cholesky decomposition of \inv{R}, so that \mat{U} is an upper triangular matrix. Then $\mat{W}_u\equiv\mat{U}$ is an upper-triangular whitener.
\item Let $\mat{R}=\mat{L}\mat{L}^T$ be the Cholesky decomposition of \mat{R}, so that \mat{L} is a lower triangular matrix. Then $\mat{W}_\ell \equiv\inv{L}$ is a lower-triangular  whitener.
\end{enumerate}
If we order the components of the data vector \vec{d} as running from earlier to later times, then this last $\mat{W}_\ell$ has the appealing property of being strictly causal, in that $\tilde{d}_i$ depends on $d_j$ only for $j\le i$.

%---------------------------------------------------------------------------------------------------------------------
\subsection{ARMA representations of the noise-covariance function}
\label{sec:arma}

Instead of pursuing superfast Toeplitz solvers for this work, we have chosen to make approximations to the noise-covariance function $r_t$ that defines \mat{R}. The extreme approximation is simply to pretend that the noise is strictly white, and therefore \mat{R} is diagonal. Though convenient, this approximation is inappropriate for typical microcalorimeter data. We have found in practice that making even an apparently very rough approximation to $r_t$ can yield performance that is much closer to the ideal than the na\"ive assumption of white noise does.

Specifically, we approximate $r_t$ as the sum of a small number $p$ of decaying exponentials. These exponentials can be complex, allowing oscillatory as well as decaying behavior in $r_t$. A simple and very widely studied class of stochastic models yields exactly this form for the noise-covariance function. In the signal-processing literature, 
these models are known as \emph{autoregressive moving-average} (ARMA) models. The theory of ARMA models is described at book-length in \citet{BoxJenkins:1994} and \citet{BrockwellDavis:2009}.  Briefly, the outlook is that some ``input'' process of independent zero-mean, unit-variance Gaussian deviates $\epsilon_i$ is passed through an infinite impulse response filter to produce the correlated noise of the ``output'' $y_i$.  When the filter consists of an order-$q$ moving average filter and an order-$p$ autoregressive filter, it is called an ARMA($p,q$) process:
\begin{align}
\lefteqn{
y_i + \phi_1y_{i-1}+\dots \phi_p y_{i-p} = } & \nonumber \\
& \hspace{1cm} \sigma^2\left[
	\epsilon_i + \theta_1\epsilon_{i-1}+\dots \theta_q \epsilon_{i-q}
	\right] \nonumber
\end{align}

A random sequence $\{y_i\}$ generated in this way has a noise-autocovariance function that obeys
\be \label{eq:arma_covar}
r_t = \sum_{j=1}^p\, a_j x_j^{|t|}, 
\mathrm{for}\ |t| \ge q-p+1,
\ee
where the exponential bases $\{x_j\}$ are the inverse of the roots of the polynomial $1+\phi_1z+\dots \phi_pz^p=0$.
Stability of the ARMA process requires that $|x_j| < 1$. For $r$ to be real, it is also required that each $a_j$ and $x_j$ be either real, or a member of a complex-conjugate pair such that $a_{j+1} = \overline{a_j}$ and $x_{j+1} = \overline{x_j}$.
We have assumed for simplicity that the roots $\{1/x_j\}$ are distinct.  In Equation~\ref{eq:arma_covar}, we see that  if $q\ge p$, the sum-of-exponentials form applies only after one or more exceptional values for small $|t|$. In the current work, we have used $q=p$ throughout, so that $r_0$ is the only exception to the general form. This choice works well when a large part of the noise variance is white, that is, concentrated in the $r_0$ component.

The appeal of approximating the noise as the result of an ARMA($p,p$) process is that its 
noise-covariance matrix \mat{R} can be factored as $\mat{R}=\mat{L}\mat{L}^T$ in $O(Np^2)$ time, 
the result can be stored in $O(Np)$ memory, and the whitener $\mat{W}_\ell=\inv{L}$ can be applied in $O(Np)$ time. This approximation thus gives us a way to evaluate the maximum-likelihood pulse amplitudes (Equation~\ref{eq:max-like}) and all other nuisance components in  a time strictly linear in the length $N$ of the data record.

It is beyond the scope of this paper to explain, in general, how to work from an observed noise covariance and choose the order $p$ of an ARMA model or its coefficients.  The work of \citet{DeGroen:1987uh} explains how one can estimate the exponential bases $\{x_j\}$ of Equation~\ref{eq:arma_covar}, though little guidance is given on the choice of the model order $p$. Once the bases are chosen, estimating the amplitudes $\{a_j\}$ is straightforward. Moving from these numbers to a fast solver of the approximate $\mat{R}$ is the topic of a future manuscript, now in preparation.  

\begin{figure}[t!]
\includegraphics[width=3.15in]{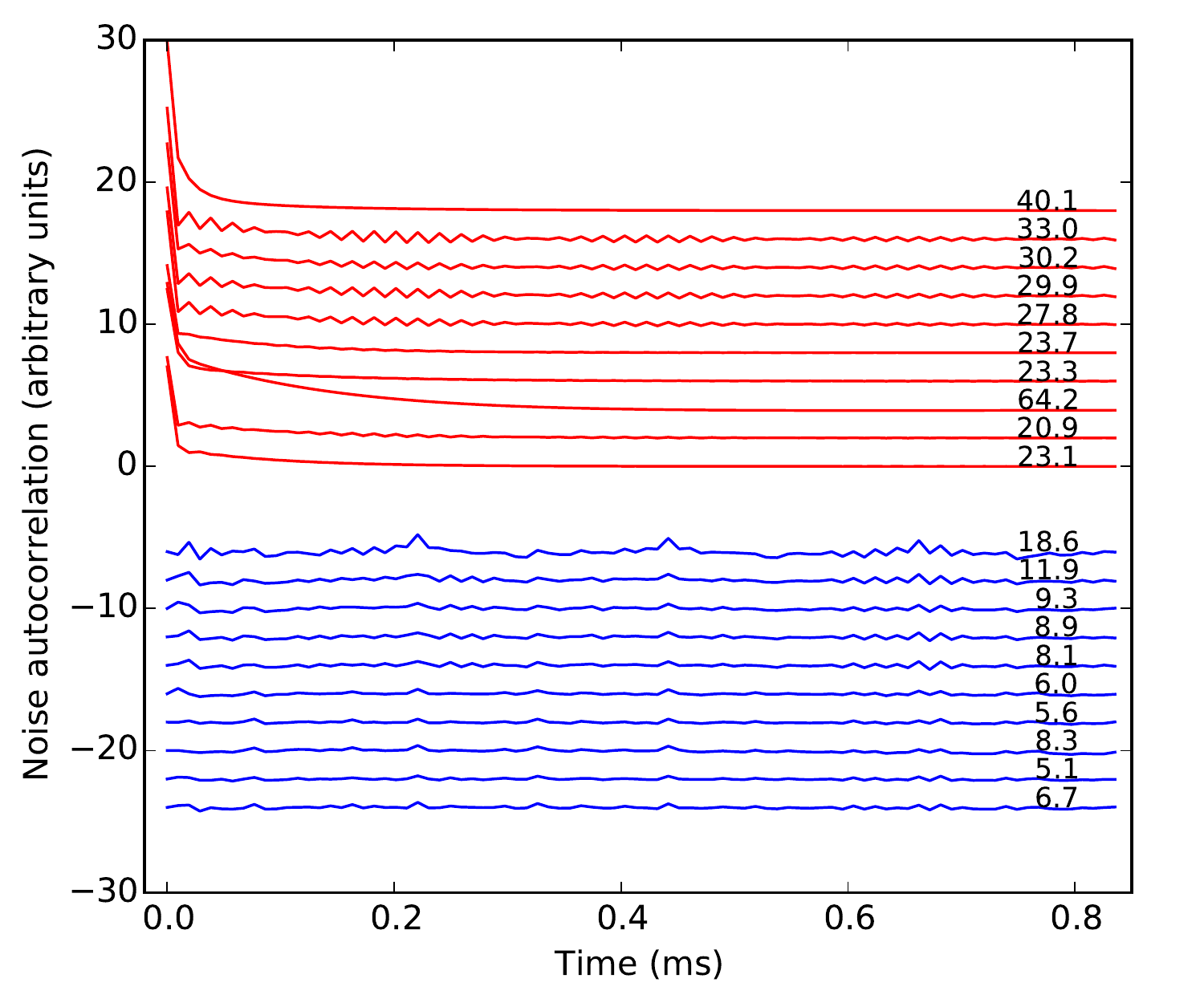}
\caption{\label{fig:arma_noise_example}
  Top ten curves (red): The ARMA(3,3) approximation to the noise covariance function for ten (out of 45) microcalorimeter detectors used in the synchrotron experiment described in Section~\ref{sec:nsls_spectrometer}.  
 % Although the units are ``arbitrary'', they are approximately equal to (eV)$^2$. 
 Bottom ten curves (blue): The difference between the noise covariance estimated from data and the ARMA(3,3) approximations.  The numbers at right give the sum of the absolute deviations of each curve from zero, showing that the residual is smaller but not dramatically smaller than the noise. Nevertheless, the approximate noise model allows for excellent \inv{R} weighting.
  } 
\end{figure}

Figure~\ref{fig:arma_noise_example} shows ten example ARMA(3,3) noise models along with the residual difference between measured and modeled noise. Although visible structure is apparent in the residuals, calculations in Section~\ref{sec:effect_noise_approx} show that the discrepancies between measurement and model are small enough to have only negligible effect on the precision of parameter estimates.

%---------------------------------------------------------------------------------------------------------------------
\subsection{Noise as an ARMA(1,1) process}
\label{sec:factoring_arma11}

In some situations, it might suffice to approximate the noise as an ARMA(1,1) process. For such cases, and to sketch the form of the solution when $p>1$, we show explicitly how to whiten data with ARMA(1,1) noise.
The result summarized here is completed in Appendix~\ref{appendix:arma11}.

We seek the whitener of type 3, the inverse of the lower Cholesky factor of \mat{R}, which is also lower triangular. First, write the noise-covariance function as
\be \label{eq:arma11}
r_t = w\delta_{t,0}+a\phi^{|t|}.
\ee
We require $|\phi|<1$, and all parameters $\{w, a, \phi\}$ to be real. The amplitude of the decaying term is $a$, and the additional white noise term is $w$.  

Let the matrix \mat{\Phi} be the banded lower-triangular Toeplitz matrix with 1 on the diagonal and $-\phi$ on the first sub-diagonal, and zeros elsewhere. The product $\mat{S}\equiv\mat{\Phi}\mat{R}\mat{\Phi}^T$ is a symmetric tridiagonal matrix, which can be factored in ${\cal O}(N)$ time as $\mat{S}=\mat{B}\mat{B}^T$.  
We can rewrite this as $\mat{R} = \inv{\Phi}\mat{B}\mat{B}^T\mat{\Phi}^{-T}$.
The desired type-3 whitener is therefore the lower-triangular matrix
\be
\label{eq:arma11_whitener}
\mat{W} \equiv \inv{B}\mat{\Phi}.
\ee
Expressions for elements of \mat{S}, recursions to compute elements of \mat{B}, and the solution to apply \mat{W} for ARMA(1,1) noise are all given in Appendix~\ref{appendix:arma11}.

We expect that the ARMA(1,1) approximation (Equation~\ref{eq:arma11}) improves on the pure white-noise approximation enough to be a valuable first step in many applications. If not, and if computational costs are not too high, multi-pulse fitting (Equation~\ref{eq:max-like}) with a direct Toeplitz solver for an arbitrary Toeplitz \mat{R} matrix is an alternative to approximating the noise as an ARMA process.

%---------------------------------------------------------------------------------------------------------------------
\subsection{Estimating the effect of approximate noise whitening}
\label{sec:effect_noise_approx}

In this section, we have proposed a fast noise whitening where the noise covariance matrix is only \emph{approximated} as \mat{R}. That is, $\mat{W}\mat{R}\mat{W}^T \approx \mat{I}$, but the equality is inexact. What is the effect of this approximation on the estimates of pulse heights and other parameters when we compute them using the approximate optimal filter given in Equation~\ref{eq:max-like}? Returning to Equation~\ref{eq:param_covar_long} for the expected parameter covariance, we see that \mat{R} arises in the middle of that equation as the covariance of the data vector: $\mathrm{E}[\vec{d}\vec{d}^T] - \mathrm{E}[\vec{d}]\mathrm{E}[\vec{d}^T]$. If we believe the \emph{true} covariance of the data to be some matrix $\mat{T}\ne\mat{R}$, then the estimate of parameter covariances  becomes
\begin{align}
\label{eq:true_Cpp}
\mat{C}_{pp} &= \inv{A}\mat{M}^T \inv{R} \mat{T} \inv{R} \mat{M} \inv{A}.
% \\ &= \inv{A}\tilde{\mat{M}}^T \mat{W}\mat{T}\mat{W}^T \tilde{\mat{M}}\inv{A}
\end{align}
This equation can be used to assess whether any particular approximation \mat{R} to the true noise covariance \mat{T} will appreciably degrade the statistical power of a multi-pulse fit or of a traditional or constrained optimal filter. For example, we can use it to evaluate what value of $p$ is sufficient when approximating the noise as an ARMA($p,p$) process. In the example analyses below, we have used Equation~\ref{eq:true_Cpp} to  determine that an ARMA(3,3) model suffices for the noise examples shown in Figure~\ref{fig:arma_noise_example}.

%=============================================================================
\section{Sub-keV x-ray fluorescence: an example application}
\label{sec:nsls_data}
%=============================================================================

We have performed measurements at a synchrotron using beam energies less than 1\,keV, establishing the viability of 
multi-pulse fitting at very high rates in the close-to-linear regime of the TES sensors\footnote{The specific TES design
employed here works up to 10\,keV, so the sensors are quite linear below 1\,keV}.  The two-peaked nitrogen 
K-line emission from NH$_4$NO$_3$ samples allows us to estimate the energy resolution at 390\,eV and to compare the 
performance of multi-pulse fitting against standard optimal filtering at that energy and a mean photon rate of 46 
cps.  Measurements of the beam scattered from a Nylon target demonstrate that good energy resolution is achieved 
even at 260 cps while retaining over 80\,\% of the photons.

%---------------------------------------------------------------------------------------------------------------------
\subsection{The NIST spectrometer at the NSLS}
\label{sec:nsls_spectrometer}

We developed a TES array of 60 sensors and installed it on the soft NIST U7A beamline at the NSLS, where it 
was in use from 2011 until the NSLS permanently ceased operation in late 2014.  The U7A beamline source
is a bending magnet with an energy range of 180 to 1200\,eV, which is optimized for operation to study the K lines of 
boron, carbon, nitrogen, 
oxygen, and fluorine; and the L lines of elements potassium to gallium. The NIST spectrometer \citep{Ullom:2014er} used three time-division multiplexer channels \citep{Reintsema:2003} each reading out twenty TES sensors apiece.  The multiplexing electronics switched 
between sensors every 640\,ns, so  the sampling rate for any given sensor was one sample per 12.8\,$\mathrm{\mu}
$s.  In the measurements described here, taken on June 14, 2012, some 40 to 45 sensors were operating.

The TES detectors used in the measurements described in this and the next section have pulses with a characteristic exponential decay time of 1.0 to 1.2\,ms. The \emph{Athena} mission's X-IFU focal plane will employ TESs approximately six to eight times faster. Therefore, pulse-analysis techniques which we show to work with the current sensors at 100 to 200 cps  per detector might scale to 1000 cps with each \emph{Athena} TES.

%---------------------------------------------------------------------------------------------------------------------
\subsection{Demonstration spectra}
\label{sec:nsls_spectra}

\begin{figure*}[ht!]
\includegraphics[width=6.5in]{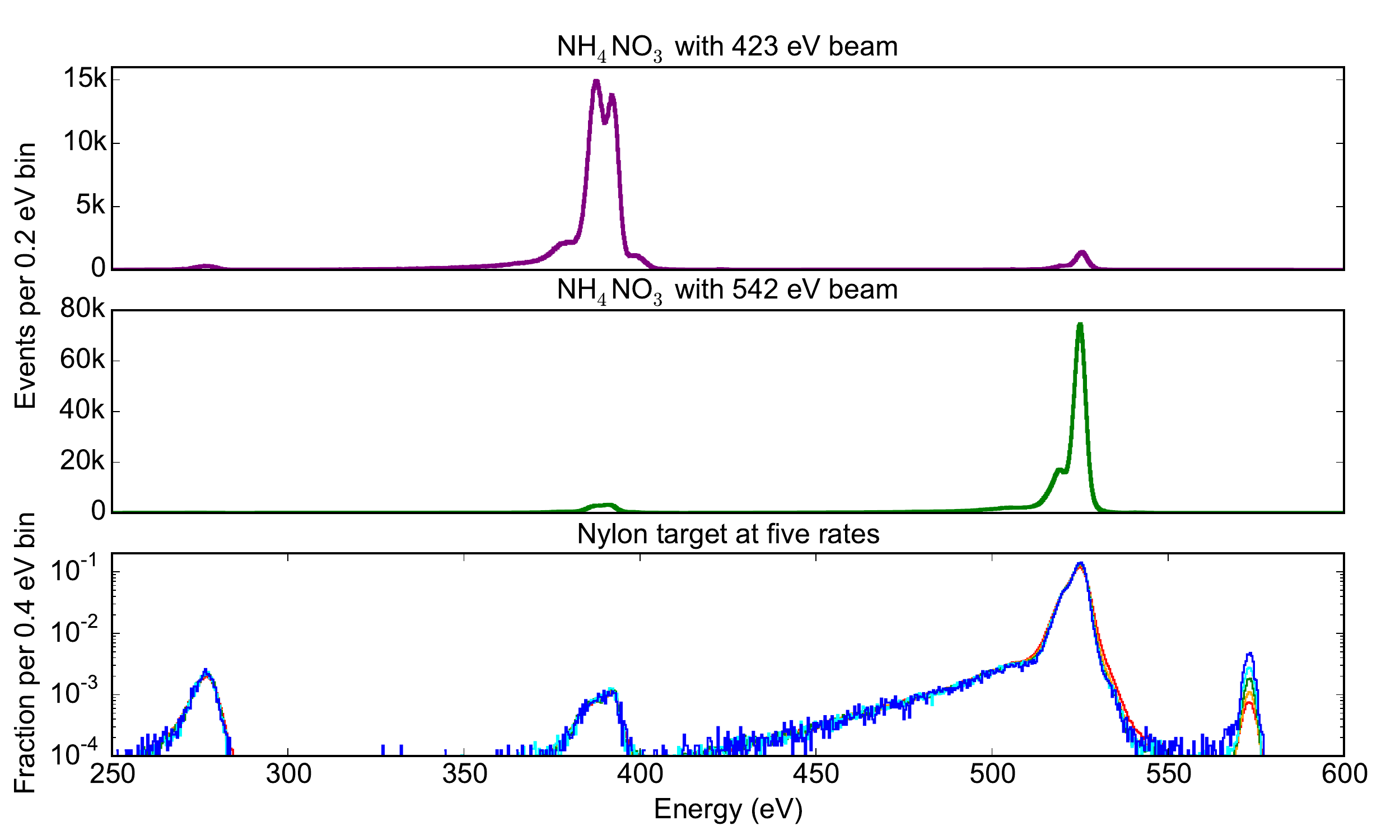}
\caption{\label{fig:wide_n_spectra}
  X-ray fluorescence spectra, the combined results from 42 sensors at the NSLS, analyzed with MPF.  In the top panel, the target is NH$_4$NO$_3$ and the beam energy is 423\,eV; most emission is in the nitrogen K-line complex ($\sim 390$\,eV). In the middle panel, the target is the same but the beam energy is 542\,eV; the overall x-ray rate is three times higher, and most emission is in the oxygen K line ($\sim 525$\,eV). In the bottom panel, the target is Nylon, and the beam energy is 573\,eV. Five spectra are shown, corresponding to a range of photon intensities that produce as few as 10 and as many as 280 counts per second per sensor.  We use the small fraction of photons elastically scattered by the target as a way to assess the energy resolution at each rate.
}
\end{figure*}

Figure~\ref{fig:wide_n_spectra} shows the x-ray fluorescence spectra from seven measurements over an energy range including the K lines of carbon, nitrogen, and oxygen.  A multi-pulse fitting analysis was used to obtain each spectrum. Two spectra show the emission from NH$_4$NO$_3$ when the probe beam was tuned to 423\,eV and to 542\,eV. In the former case, the nitrogen emission ``line'' is split into a two-peaked complex. Chemical effects cause the splitting, which results from the very different bonding environments of the nitrogen atoms in the ammonium and in the nitrate ions \citep{Vila:2011jj}. The split nitrogen line offers an excellent opportunity to assess the energy resolution resulting from different analysis methods.
\begin{figure*}[ht!]
\includegraphics[width=\textwidth]{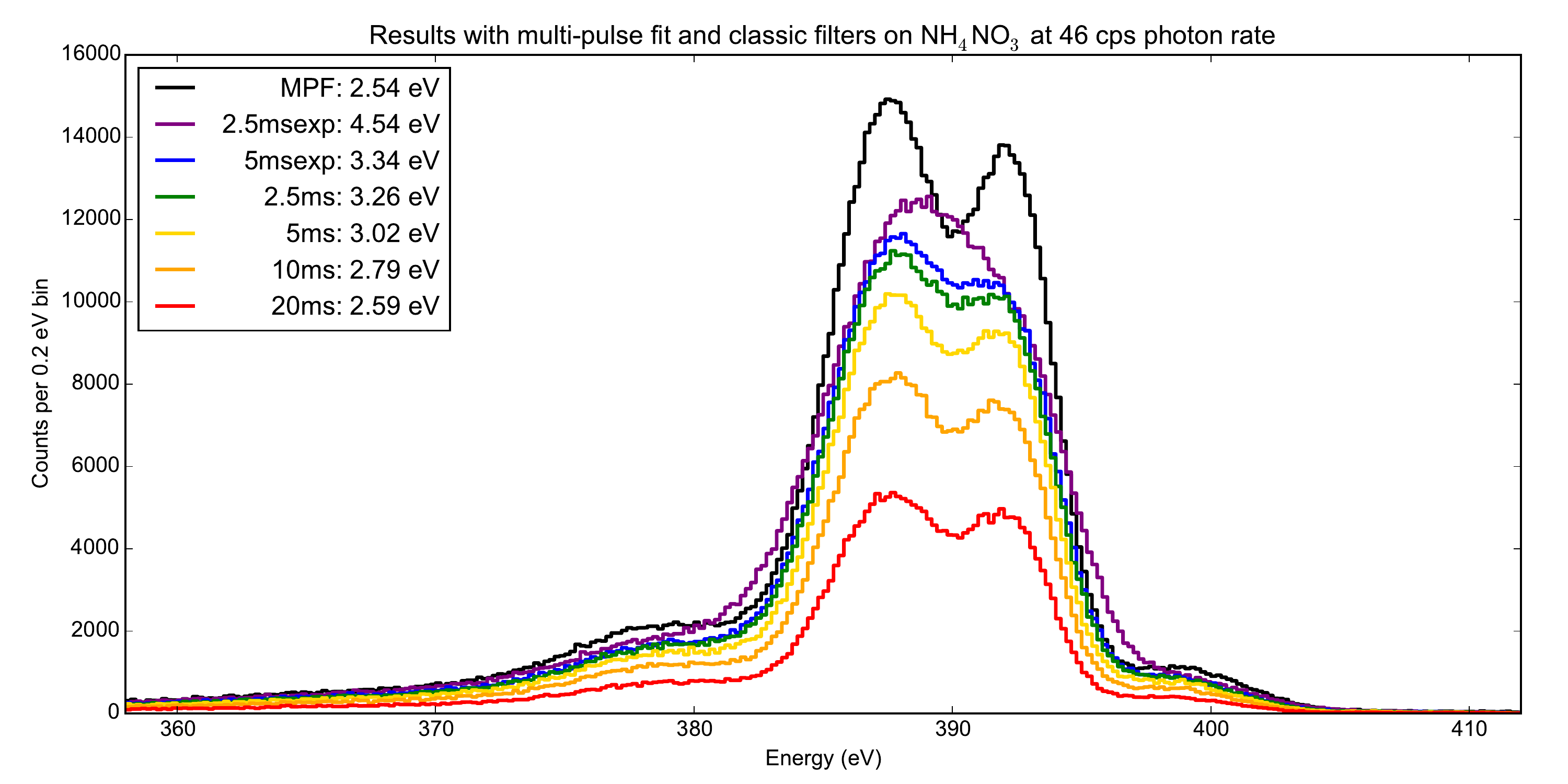}
\caption{\label{fig:nh4no3_spectra}
  The two-peaked nitrogen-K x-ray fluorescence line emitted by a NH$_4$NO$_3$ target illuminated by 423\,eV probe beam.  The seven spectra correspond to the same 1,038,000 photons analyzed in seven different ways. The six lower spectra were generated by standard optimal filter analysis, with six different choices of filter. The top spectrum is the result of multi-pulse fitting of the same data. The legend gives the estimated energy resolution (FWHM of Gaussian broadening). The MPF analysis achieves energy resolution as good as the longest standard optimal filter's, while also making use of more pulses than even the shortest, lowest-resolution constrained optimal filter. The vertical scale is the same for all spectra---taller 
  spectra have more usable counts.
}
\end{figure*}

To achieve the best possible energy resolution, a few small corrections are required after the basic multi-pulse fit described in Section~\ref{sec:fitting_multiple}. One is a small gain-drift correction. The gain of each TES sensor  is found to be anticorrelated with its steady-state baseline level. Although each gain varies by only a few parts per thousand, the derived energy resolution improves somewhat when fitted pulse heights are scaled by a number linear in the baseline level. The correction has a single variable parameter per sensor, which is chosen by minimization of the Shannon entropy of the corrected pulse-height spectrum. Lower entropy corresponds to our general sense of  ``sharper'' spectral features, so this heuristic is appropriate in a spectrum dominated by unresolved or marginally resolved lines.

A similar one-parameter correction is applied to remove a small, quadratic dependence of pulse height on the pulse arrival time. (The arrival time is found via a first-order linear model for the effect of time offset, as described in Section~\ref{sec:arrival_time}.) Again the parameter is chosen to minimize the spectral entropy. Both the gain-drift and the arrival-time corrections are generally employed in our standard optimal-filter analysis, and are not unique to the multi-pulse fitting technique.

The third correction is unique to multi-pulse fitting, but only because this technique permits analysis of pulses with far more pileup than is possible in other analysis approaches. We make a small correction quadratic in the ``residual current'' flowing through the sensor just before the present pulse arrives. As with the others, this correction requires only one parameter per sensor, is not more than a few eV in the most extreme cases, and is selected by entropy minimization of the corrected spectrum.  At low pulse rates, this correction has no measurable effect on the data quality, but at the highest rates, it can improve the resolution from 5 or 6\,eV to 4\,eV.

Multi-pulse fitting is intended as a method to improve on the either the photon efficiency or the energy resolution (or both) offered by standard optimal filtering. Using the NH$_4$NO$_3$ data with 423\,eV x-ray beam (Figure~\ref{fig:wide_n_spectra}, top panel), we have made a direct comparison of MPF and standard filtering. The per-detector average photon rate in these data is 46\,cps. We compare the MPF results against the results from a suite of six different optimal filters. Four are of the standard type, in which the filter is explicitly insensitive only to addition of a constant level. These filters are 20, 10, 5, and 2.5\,ms long (the shortest is 195 samples long, given the 12.8\,$\mu$s sample time).  The other two filters are constrained optimal filters (Section~\ref{sec:opt_filtering}) of length 5 and 2.5\,ms, with the constraint being that they are insensitive not only to the addition of a constant, but also to decaying exponentials with time constants 0.64 and 1.024\,ms (50 and 80 samples). These time constants correspond to the typical sensor's signal decay at a few ms and at a few tens of ms after the peak of a pulse. Multiple decay time constants are not unusual in microcalorimeter pulses.

Each of the six standard or constrained optimal filters has a corresponding ``veto window''---effectively
a dead time imposed offline---inside of which no other pulse is permitted. Each filter is therefore applied to a unique subset of the full 1,038,000 pulses observed in this measurement. For the four standard filters, it was required that the previous pulse arrive at least 4\,ms before the start of the filtered record, and that the next arrive at least 0.1\,ms after the end of the record. The veto windows are thus 24.1, 14.1, 9.1, and 6.6\,ms long.  As the constrained filters are designed specifically for insensitivity to the tails of any prior pulses, their requirement was relaxed so that the prior pulse could arrive as little as 0.5\,ms before the start of the filter. This yields windows of length 5.6 and 3.1 ms. At the mean data rate of 46\,cps, the expected efficiency of the timing cuts ranges from 33\,\% for the longest standard filter to 87\,\% for the shorter of the constrained filters.

Figure~\ref{fig:nh4no3_spectra} shows spectra generated with the six optimal filters and with MPF.  As described above, the optimal filters can increase photon efficiency by employing shorter data records, by building in insensitivity to the exponential tails of previous pulses, or both. The increased efficiency always comes at the price of poorer energy resolution, as shown by the family of spectra; the more photons in a spectrum, the less distinct the two nitrogen fluorescence peaks become.   The MPF results, on the other hand, demonstrate energy resolution as good as the longest standard filter's while achieving 96\,\% photon throughput, well above that permitted by any optimal filter's corresponding timing cuts. 

%---------------------------------------------------------------------------------------------------------------------
\subsection{Energy resolution at various count rates with MPF}
\label{sec:nylon_resolution}

To understand how the energy resolution changes with the photon rate, we use the measurements made of fluorescence and scattered x-rays from the Nylon target with a 573\,eV beam (Figure~\ref{fig:wide_n_spectra}, bottom panel). For this purpose, we study the width of the peak corresponding to elastically scattered beam photons. They make up only some 1\,\% of the total photons, but their value is that the natural line width (set by the beamline's monochromator) is approximately 0.6\,eV FWHM, which is too narrow to be resolved with the TES spectrometer. Therefore, the measured peak width serves as a direct measurement of the instrumental resolution. 

We varied the intensity of the scattered and fluorescence x-rays by moving the spectrometer. The detector array was placed approximately 2\,cm from the Nylon target to achieve the highest intensity and 11\,cm for the lowest intensity. At the higher rates, the sensors in one of the three multiplexer columns were much closer to the fluorescence-emission spot on the target, and the x-ray pulse rate was much higher for these sensors. Consequently, we treat the five spectrometer positions as measurements at \emph{ten} distinct data rates, handling the higher-rate column (18 of the TESs) separately from the other $\sim24$ sensors.

\begin{figure}[t]
\includegraphics[width=.48\textwidth]{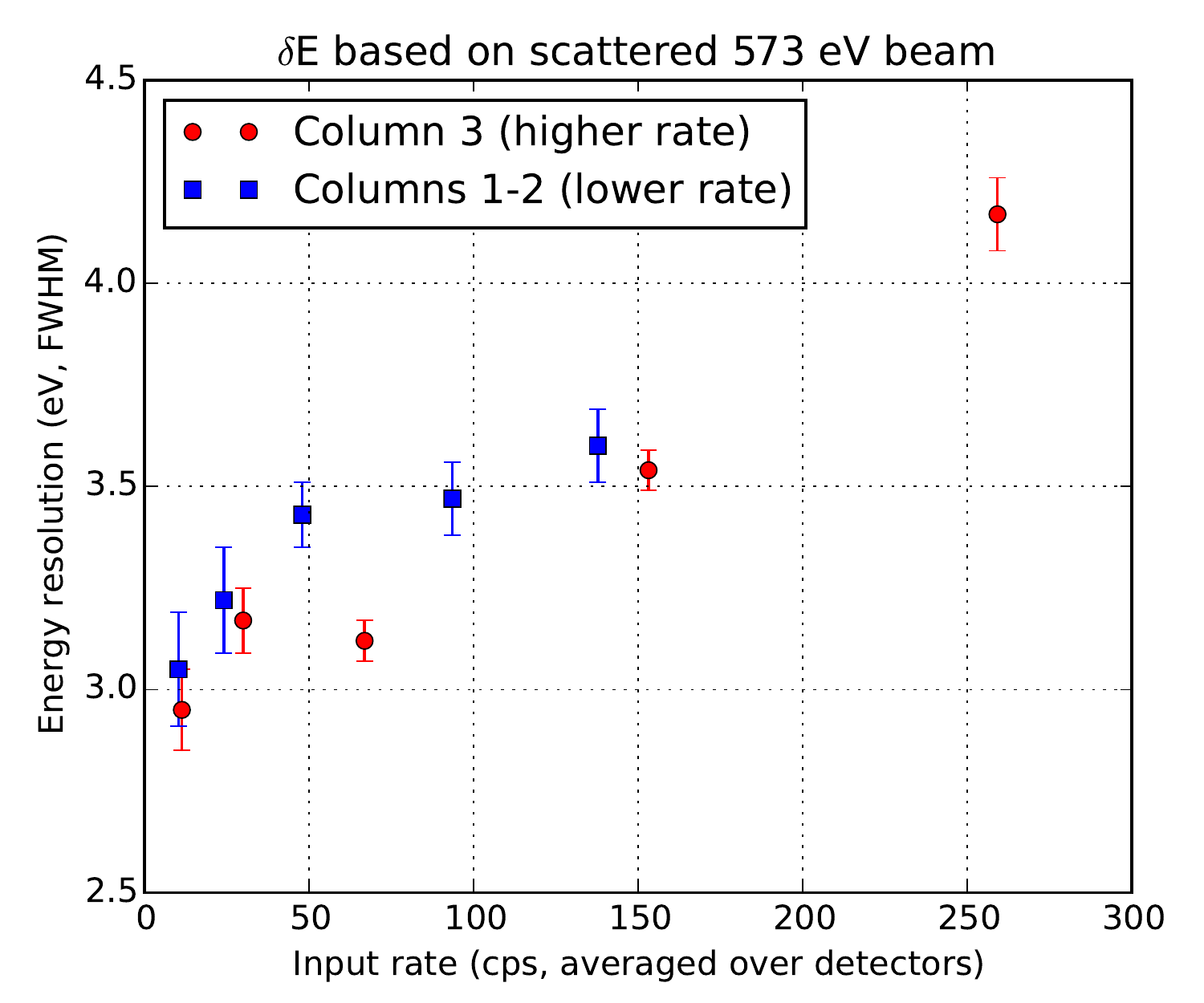}
\caption{\label{fig:nylon_res_vs_rate}
  Estimated energy resolution at 573\,eV based on multi-pulse fitting analysis of the Nylon target data, as a function of the incident photon rate (per sensor). The energy resolution reported is the estimated full-width at half-maximum of the best-fit Gaussian to the scattered-beam peak. The fit is performed with photons between 568\,eV and 578\,eV; a line is included in the fit to allow for background photons.    Multiplexer Column 3 is handled separately from the others, because---for geometric reasons---the photon rates observed in Column 3 are substantially higher.  The resolution degrades only from 3.0 to 4.2\,eV at 262 cps, even though multi-pulse fitting accepts 80\,\% of all photons at that rate.
  }
\end{figure}

We estimate the spectrometer resolution by fitting a Gaussian peak (plus a line to represent background photons), always using data from 568 to 578\,eV. The parameters are chosen to maximize the full Poisson likelihood, in order to reduce the expected parameter bias \citep{Fowler:2014JLTP}. Figure~\ref{fig:nylon_res_vs_rate} shows the results. The resolution is found to degrade only a small amount even as the x-ray count rate is increased from 10 to 262 cps.  Pulses are cut only if the prior pulse falls within 0.55\,ms or the next pulse falls within 0.32\,ms.
 At the highest rate, 80\,\% of the pulses survive these cuts, yielding a 4.2\,eV resolution at an input rate of 260 cps per sensor.

The bright oxygen K line does not have a known line shape that we can use as a primary tool for estimating energy resolution. Nevertheless, we can use the line to corroborate the resolution found using the scattered beam. A Kolmogorov-Smirnov test~\citep{NumRec3:2007} shows complete consistency between the O line shapes for the two lowest-rate measurements (roughly 10 and 25 cps), so we combine these and assume them to be the results of smearing the true shape by some unknown amount. The higher-rate data are then fit to find the \emph{additional} Gaussian smearing required to make the best agreement.  If we assume that the resolution of the two lowest-rate data sets is 3.0\,eV FWHM, as established from the scattered beam, then the total resolution implied by the oxygen line shape is also consistent with the scattered beam results at higher photon rates: 3.6\,eV at 150 cps and 4.5\,eV at 260 cps.

%---------------------------------------------------------------------------------------------------------------------
%\subsection{The tradeoff of photon efficiency against energy resolution }
%\label{sec:nsls_efficiency}

\begin{figure}[th!]
\includegraphics[width=.48\textwidth]{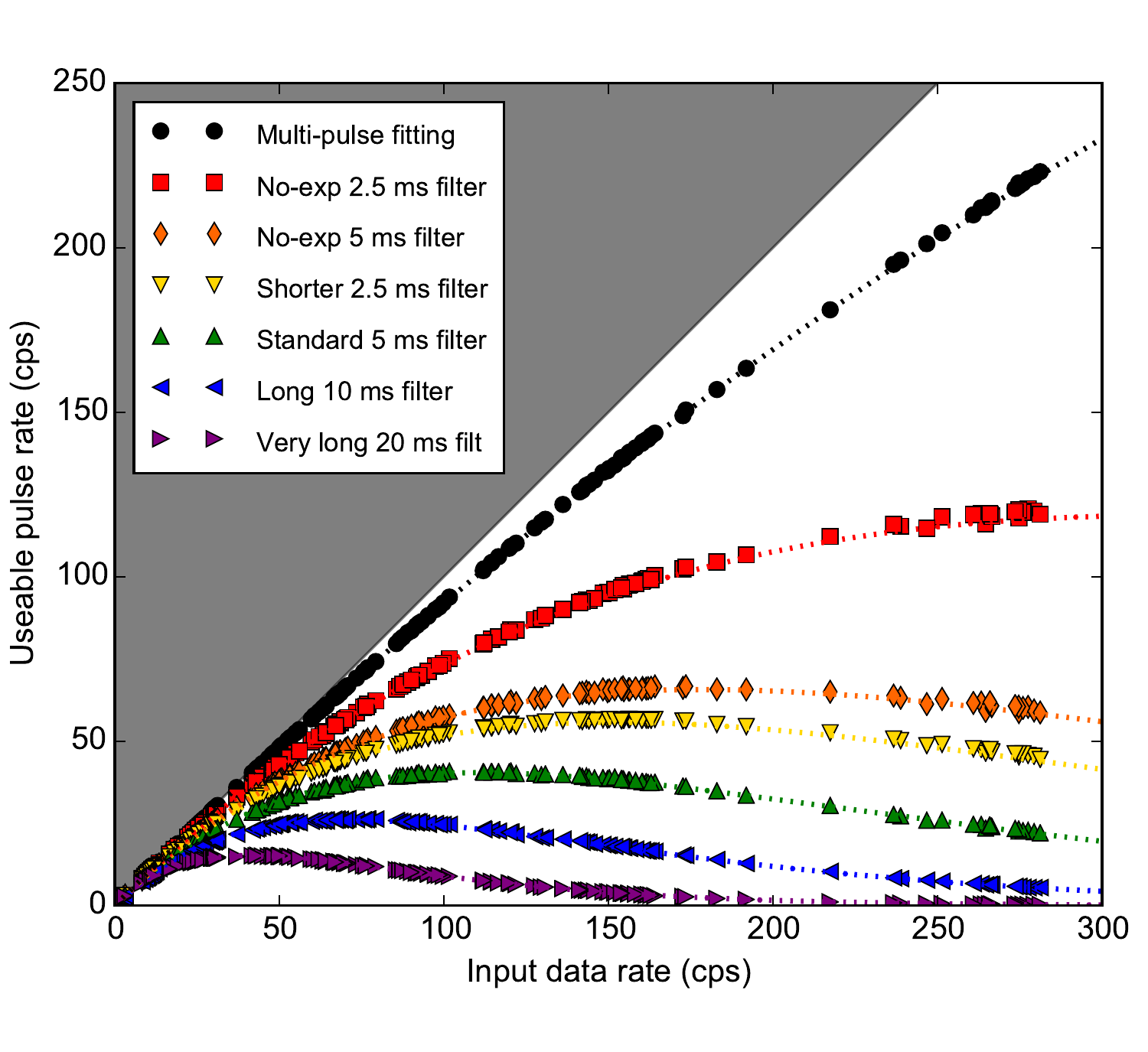}
\caption{\label{fig:ratesvsrates_nylon}
  The per-sensor rate of usable photons in the nylon data as a function of the total rate of photons. Each color and marker shape corresponds to a different analysis. The circles represent the multi-pulse fitting. There are approximately 200 circles: one for each of the 40 working TESs at each of five different fluorescence photon rates. The other six marker shapes correspond to a traditional analysis using optimal filters of various fixed lengths. Each of these analyses has a corresponding ``veto window'': a pulse has to be cut if another photon arrives during its veto window. The lines represent the theoretical usable rate $r\mathrm{e}^{-rw}$, where $r$ is the raw photon rate and $w$ is the duration of the relevant veto window ($r$ and $w$ are given, not fit).  The gray shading (upper left) indicates the disallowed region, where more than 100\% of pulses would be usable.
  }
\end{figure}

Figure~\ref{fig:ratesvsrates_nylon} shows the rate of photons passing all timing and other cuts as a function of the incident photon rate. Each sensor appears seven times, once for the MPF analysis and for each of six standard optimal filter analyses. Although cuts other than timing are applied, the throughput is largely governed by timing cuts, and so the measured data all lie close to the theoretical lines.

In summary, the Nylon fluorescence results show that $\delta E$ of approximately 4\,eV is possible at rates exceeding 250 cps, a resolution consistent with that provided by a traditional filter of length 10\,ms. Yet the MPF can achieve this using a vastly higher fraction of the photons than the traditional filter can.
Furthermore, we have established that the technique is workable with dozens of separate detectors and over a very wide range of photon rates.

%%---------------------------------------------------------------------------------------------------------------------
%\subsection{}
%\label{sec:}

%=============================================================================
\section{Multi-pulse fitting in the nonlinear regime: a second example}
\label{sec:mn_data}
%=============================================================================

To complement the low-energy limit, discussed in the previous section, we also made measurements at 5900\,eV. These data consist of the K$\alpha$-line fluorescence emission of manganese metal. For this purpose, electrons are accelerated in a standard commercial x-ray tube source onto a rhodium target, which emits brehmsstrahlung photons that illuminate the manganese secondary target. By varying the electron current, we are able to produce x-ray count rates on a TES microcalorimeter up to hundreds of counts per second. We performed the measurement using a single TES in a non-standard non-multiplexed readout system. In this system, the single detector's sample rate was one sample per 640\,ns. Examples of the raw TES current appear in Figure~\ref{fig:example_highrate}.

% Corrections and res assessment
Multi-pulse fitting was performed on the manganese fluorescence data in the same manner described in the previous section. The same small corrections detailed in Section~\ref{sec:nsls_spectra} were applied here: gain-drift, arrival-time, and a leading-order nonlinearity correction. The sensor resolution was assessed via the two-peaked shape of the Mn K$\alpha$ line complex, assuming the natural line shape given by \citet{Holzer:1997ts} as a sum of seven Lorentzian profiles. 

\begin{figure}[t!]
\includegraphics[width=.48\textwidth]{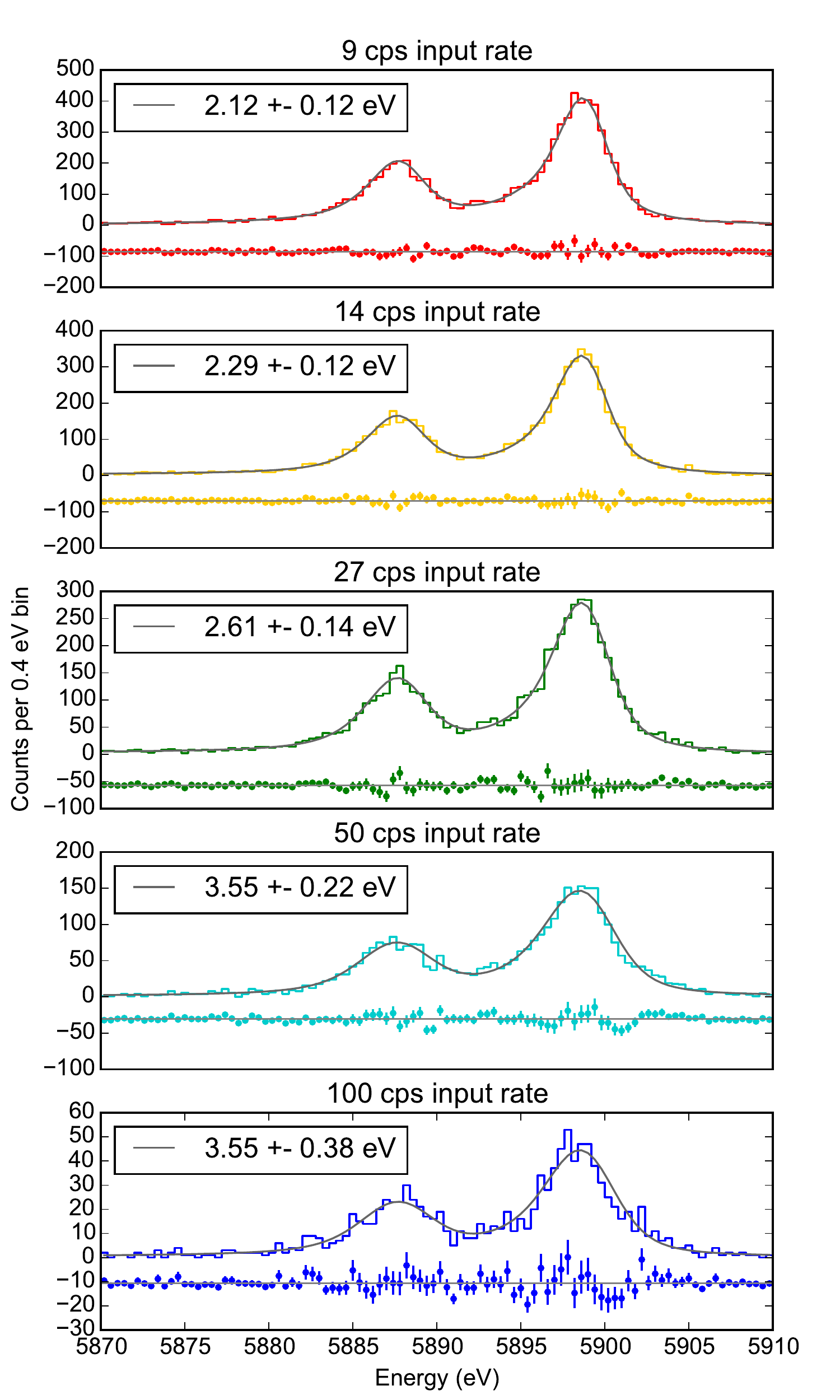}
\caption{\label{fig:mnka_spectra}
The Mn K$\alpha$ fluorescence-line spectrum, determined through multi-pulse fitting, 
at five input photon rates.
These spectra correspond to the strictest of the three time-isolation cuts, as shown in 
Table~\ref{tab:mnka_results}.  In each section of the figure, the measured spectrum is shown as a histogram, the best-fit curve is the smooth gray curve, and the data-minus-fit residuals are shown (offset for clarity) as points with error bars extending to $\pm\sqrt{N}$ to indicate the approximate range of consistency with the Poisson distribution of counts in each bin.}
\end{figure}

The measured energy spectrum at 6\,keV is not simply a convolution of the ideal energy spectrum with a Gaussian, as we could safely assume in the previous section.  Observations indicate that an exponentially decaying tail towards low energies is an important component of the energy response.
Such a tail is discussed quantitatively in \citet{Bortels:1987}, for example. In most Mn K$\alpha$ data, we find that approximately 10\,\% of all pulses must be attributed to the exponential tail, which has a typical scale length of 20 to 30\,eV\@.  Both the fraction and the scale length are allowed to vary when fitting for the instrument response. The remaining 90\,\% of pulses are unaffected by the exponential tail. The distribution of all pulses---tail or otherwise---is then convolved with a Gaussian, and it is the full width at half-maximum of this Gaussian which we label ``the energy resolution.''  When the fraction of pulses in the tail exceeds about 20\,\%, however, it is not at all clear that the Gaussian width still represents the quantity of interest.  Whether the low-energy tail represents physical effects in the detector or systematic effects of the data analysis is unknown, but we expect that both causes play important roles. We fit for the low-energy component in analyzing all Mn K$\alpha$ spectra and indicate cases where it appears to comprise at least 20\,\% of all pulses.

%    In [124]: reload(m2011); outrate,res=m2011.many_fits(all)
%    7.18 Hz  83.85%     2.12 eV tail:  7.2% T: 10.4 eV  N=11291
%    7.78 Hz  90.84%     2.25 eV tail:  8.4% T: 16.2 eV  N=12231
%    8.35 Hz  97.51%     2.35 eV tail:  8.5% T: 11.0 eV  N=13130
%    
%    10.92 Hz  75.62%     2.28 eV tail:  8.3% T: 40.4 eV  N= 9158
%    12.41 Hz  85.96%     2.43 eV tail:  9.1% T: 29.0 eV  N=10411
%    13.96 Hz  96.72%     2.83 eV tail:  9.4% T: 23.2 eV  N=11714
%    
%    16.31 Hz  60.78%     2.61 eV tail: 21.3% T: 207.4 eV  N= 8207
%    20.44 Hz  76.18%     2.94 eV tail:  8.4% T: 20.3 eV  N=10286
%    25.21 Hz  93.98%     3.61 eV tail:  9.7% T: 25.7 eV  N=12689
%    
%    19.99 Hz  39.65%     3.55 eV tail:  2.1% T: 11.7 eV  N= 5030
%    30.51 Hz  60.51%     3.75 eV tail: 17.6% T:  8.9 eV  N= 7677
%    44.89 Hz  89.03%     4.53 eV tail: 20.4% T:  9.6 eV  N=11295
%    
%    14.67 Hz  14.71%     3.55 eV tail:  5.2% T: 14.3 eV  N= 1495
%    36.50 Hz  36.60%     4.16 eV tail: 40.7% T: 12.6 eV  N= 3719
%    78.68 Hz  78.90%     4.88 eV tail: 49.2% T: 14.5 eV  N= 8016

\newcommand{\hs}{\hspace{2em}}
\begin{table*}
\rotate
\begin{center}
\begin{tabular}{lrrrrr}
\tableline
Input rate (cps) & 9 & 14 & 27 & 50 &\  100\ cps \\ \tableline
MPF pulse rates: \\
\hs Loose & 8.4 & 14.0 & 25.2 & 44.9 & 78.7\ cps\\
\hs Medium & 7.8 & 12.4 & 20.4 & 30.5 & 36.5\ cps\\
\hs Strict & 7.2 & 10.9 & 16.3 & 20.0 & 14.7\ cps \\
Energy Resolution: & \\
\hs Loose    & 2.35 & 2.83 & 3.61 & 4.53 & \emph{4.88}\ eV \\ 
\hs Medium & 2.25 & 2.43 & 2.94 & 3.75 & \emph{4.16}\ eV \\
\hs Strict      & 2.12 & 2.28 & 2.61 & 3.55 & 3.55\ eV  \\
\tableline
OptFilt pulse rates: & \\
\hs 2.5 ms, $\perp$ exp & 8.0 & 13.2 & 23.3 & 40.1 & 68.3\ cps \\
\hs 5 ms, $\perp$ exp &  7.9 & 12.9 & 22.2 & 36.6 & 56.4\ cps \\
\hs 2.5 ms & 7.7 & 12.2 & 20.4 & 31.6 & 41.2\ cps\\
\hs 5 ms & 7.5 & 11.9 & 19.3 & 28.0 & 32.4\ cps \\
\hs 10 ms & 7.3 & 11.3 & 17.3 & 23.0 & 21.9\ cps \\
\hs 20 ms & 6.8 & 9.9 & 13.7 & 14.9 & 9.1\ cps \\
Energy Resolution: & \\
\hs 2.5 ms, $\perp$ exp & 3.93 & 4.07 & 4.82 & 6.13 & \emph{6.41}\ eV \\
\hs 5 ms, $\perp$ exp & 3.21 & 3.36 & 4.02 & 5.43 & \emph{5.56}\ eV \\
\hs 2.5 ms & 2.95 & 3.18 & 3.40 & 4.56 & \emph{4.52}\ eV \\
\hs 5 ms & 2.82 & 3.10 & 3.32 & 4.50 & \emph{4.27}\ eV \\
\hs 10 ms & 2.58 & 2.75 & 3.20 & 3.89 & \emph{4.54}\ eV \\
\hs 20 ms & 2.45 & 2.59 & 3.02 & 3.77 & \emph{5.00}\ eV \\
\tableline
\end{tabular}
\end{center}
\caption{\label{tab:mnka_results}
The rates of usable pulses and the energy resolutions (FWHM of the Gaussian component of the energy spread function) for Mn K$\alpha$ fluorescence x-rays generated at five rates from 9 to 100 counts per second on one TES detector. The upper section shows the results from multi-pulse fitting with pulse selections from least to more restrictive. The lower section shows the results from standard optimal filtering with six different choices of filter length and constraints. Longer filters have improved energy resolution but more restrictive requirements concerning pulse pile-up.
An energy resolution in italics indicates that over 20\,\% of the data falls in the low-energy exponential tail, so the interpretation of the Gaussian width as ``energy resolution'' is suspect.
}
\end{table*}

% Res results

Figure~\ref{fig:mnka_spectra} shows the line shape and the best-fit energy resolution (Gaussian FWHM) at five different photon rates up to 100\,cps, after we performed multi-pulse fitting and imposed a strict cut primarily against pulses that are too closely piled up on the tail of the prior pulse. This cut is chosen to achieve the best possible energy resolution, though it passes only 40\,\% of pulses in the 50\,cps data set. To show that much higher photon efficiency is possible at only modest cost in energy resolution, we also apply two other cuts that are less selective against pulse pile-up.
The upper half of Table~\ref{tab:mnka_results} states the resolutions and usable pulse rate for each input rate and for the three cuts. 
The data at the highest rate of 100\,cps fit the model with a large fraction of events (40\,\% to 50\,\%) attributed to the low-energy exponential tail with either of the two less restrictive cuts.
For this reason, we are reluctant to call the width of the Gaussian component in the 100\,cps data its ``energy resolution''.

\begin{figure*}[ht]
\includegraphics[width=\textwidth]{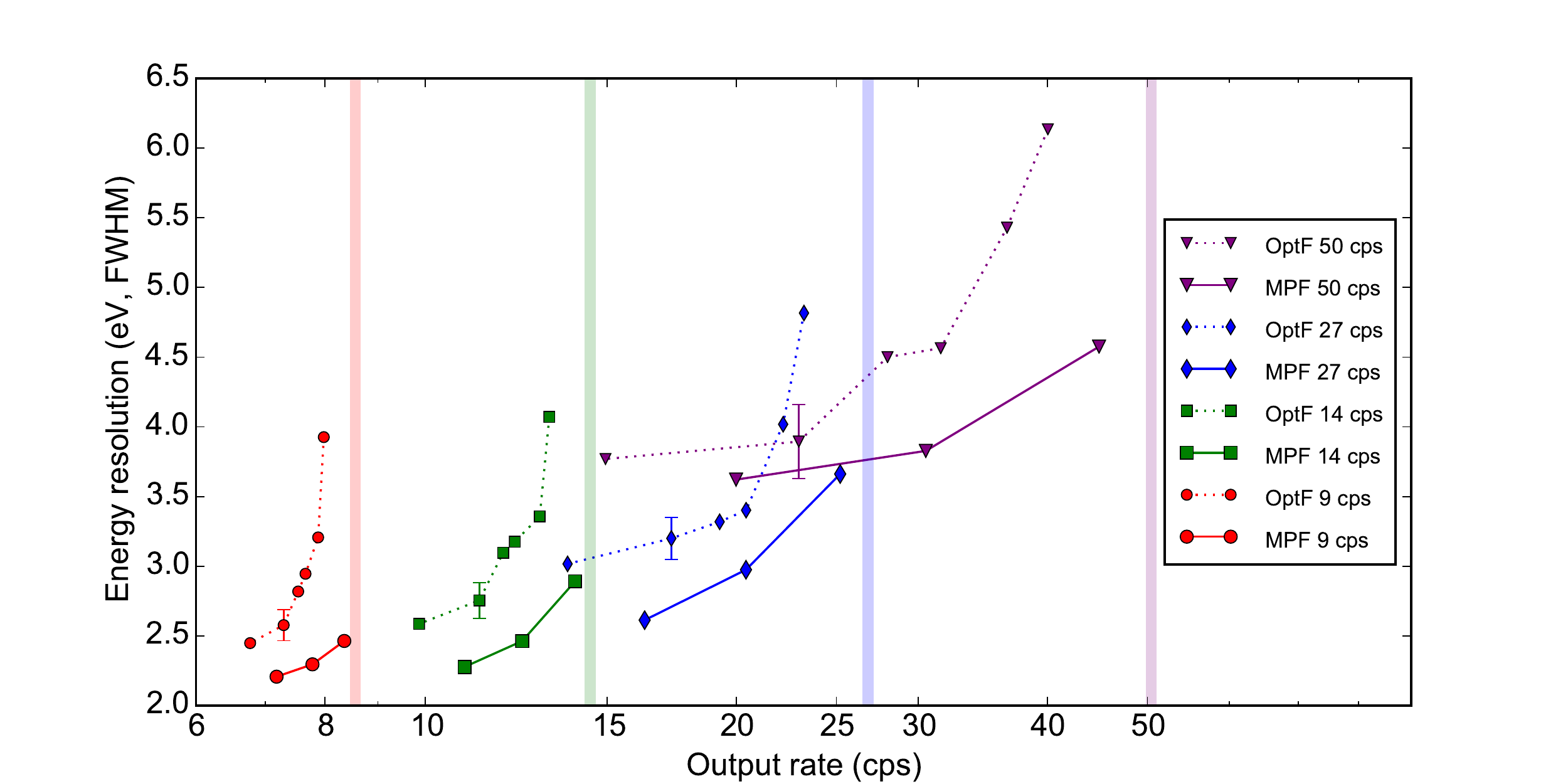}
\caption{\label{fig:mnka_res_vs_rate}
  Energy resolution determined at the Mn K$\alpha$ line complex. Incident photon rates of 9, 14, 27, and 50 counts per second are shown in different colors and with different symbols. The vertical shaded areas show the incident photon rates, which would indicate 100\,\% efficiency.  The upper curves connected by dashed lines show the classic optimal filtering analysis; each point corresponds to a different filter and its appropriate pulse cuts. The lower curves connected by solid lines show the superior results from multi-pulse fitting to the same data. The three points correspond to different selection cuts from most to least restrictive.  The 100 cps data are omitted, because they are not well fit by a purely Gaussian resolution function. (The Gaussian component of their response is typically between 4 and 5.5\,eV FWHM and is also superior for MPF.) To minimize visual distractions, only a single, representative error bar is shown for each photon rate; it reflects uncertainty in the fit between measured and modeled spectra.}
\end{figure*}

% Discuss figure 9
The usable subset of Mn K$\alpha$ pulses can be selected either with aggressive cuts to minimize systematics (and optimize energy resolution) or with more relaxed cuts to favor the highest possible pulse efficiency. In the points connected by solid lines in Figure~\ref{fig:mnka_res_vs_rate}, we show the sort of balance this allows; the points from left to right show the energy resolution as cuts vary from stricter to more permissive. The main variable among the three choices of cut criteria is how much ``leftover energy'' is allowed in the TES when a given pulse arrives---that is, how much prior signal a pulse can be piled upon. Pulses suffering the most from nonlinear detector effects can thus be eliminated.   A similar resolution-efficiency tradeoff is also possible in standard optimal filtering (points on the same plot connected by dotted lines). Here, six different filters are chosen, having the same record lengths, pre-trigger fractions, and constraints as described in Section~\ref{sec:nsls_spectra}.  

% Summarize
The multi-pulse fitting technique offers both a higher maximum possible pulse efficiency than optimal filtering \emph{and} superior energy resolution at any given level of throughput. It accomplishes this because the assumption of pulse linearity allows us to relax the pulse-isolation requirements. Analysis of any given pulse uses more data than would be possible with standard optimal filtering. This matters because pulse heights are meaningful only as a height \emph{relative to a varying baseline level}. One way to view the excellent performance of multi-pulse fitting is that one can use the extra data to make superior estimates of this baseline.

The data at the lowest rates establish that this sensor is capable of 2.1\,eV resolution at 5900 eV. The resolution degrades as the photon rate is increased. Importantly, the degradation appears equally in the multi-pulse fitting and in the optimally filtered data. It appears to be a nonlinearity effect, inadequately addressed by the simple leading-order correction. This degradation with increased rate might seem to be a weakness of multi-pulse fitting, but we do not believe that it is intrinsic to the technique. More sophisticated methods for handling the nonlinear evolution of pulse shapes and sizes with increasing ``residual sensor energy'' are needed, but they are not incompatible with the basic concept of fitting the data records as a sum of modeled components.  Even without them, MPF achieves energy resolutions of 4\,eV at 50 to 100\,cps with a slow ($\tau=1.2$\,ms) TES detector.

%=============================================================================
\section{Prospects and possible extensions}
\label{sec:prospects}
%=============================================================================

%Prospects:
The prospects for the use of multi-pulse fitting in future x-ray satellite missions depend on controlling both the computational costs of the method and the systematic errors that arise because the technique assumes linearity from an inherently non-linear sensor technology.  

%Costs
We have argued that when one approximates the noise as an ARMA process, the cost of multi-pulse fitting to a data set of length $N$ scales linearly in $N$, just as with optimal filtering. Though true, this claim ignores two problems: first, that the overall scale factor grows rapidly with the number of pulses being fit at once. If $m$ pulses are fit in the typical data section, and $n_p$ parameters are fit per pulse,% (we have used $n_p=2$ in the present work)
then multi-pulse fitting requires, among other computations, the inversion of a square matrix of size $m n_p$. This cost can grow quickly, so multi-pulse fitting should be done on the shortest reasonable segments of data. Future work will investigate alternatives for selecting data segment lengths, in order to control the number of floating-point operations without compromising energy resolution.
%Future work will investigate whether ``single-pulse fitting''---breaking up the data into sections of variable length containing exactly one pulse---is a viable alternative that could prevent the number of floating-point operations per pulse from scaling as a large power of the pulse rate.

Another costly feature is that multi-pulse fitting requires computations to be done on all samples in a long data sequence, instead of creating fixed-length records from only a fraction of the data. In the limit of high pulse rates, though, most data samples are incorporated into the fixed-length records anyway, and the difference becomes smaller.\footnote{Indeed, if records were allowed to overlap, then the optimal filter technique would need to use substantially \emph{more} than 100\,\% of the samples at high rates.}
Furthermore, even with optimal filtering of short records, one must convolve the complete raw data stream with some kernel in order to identify trigger points.

%Nonlinear data
Aside from practical concerns of computational costs, sensor nonlinearity is the most likely source of trouble with the fully linear multi-pulse fitting technique as presented here. How to address nonlinearity in this context---or in the more familiar context of optimal filtering---remains a major open issue for the entire field of microcalorimetery. We are not prepared to settle the question now, but one approach is to begin by expressing pulse records as linear combinations of basis vectors. Such a basis might be found empirically, through Principal Component Analysis of data vectors, or by imposing an \emph{a priori} model. Either way, multi-pulse fitting means precisely the approximation of pulse records as a linear combination of specified vectors; it is therefore fully compatible with a such a decomposition. Although a complete nonlinearity-aware analysis may estimate pulse energies as some \emph{nonlinear} function of the amplitudes in the decomposition, one can still benefit from the fitting algorithm we have described in this work to perform the \emph{linear} decomposition in the first place.

%Sum-of-exponentials representation?
One possible approach to accelerate further computation is to express basis vectors (after noise-whitening) as the sum of a small number of decaying (potentially complex) exponentials. If this can be done for all basis vectors, then inner products of the form $\tilde{\mat{C}}^T\tilde{\mat{C}}$ can be computed analytically, in a time independent of the data length (the height of the \mat{C} matrix). How amenable realistic basis sets are to such an expansion is unknown and is likely to vary considerably over the range of TES designs, noise environments, and spectrometer applications.

%Nonstationary noise
Another extension to the basic MPF approach that might be useful in some instances would be handling of non-stationary noise. In full generality, non-stationary noise would defeat the simplifications that the pure ARMA noise model introduces. Still, it is possible to combine the ARMA model with certain forms of nonstationarity and retain the former's speed advantages. One such form is the case of a noise-covariance matrix $\mat{R}=\mat{D}\mat{R}_0\mat{D}$, where \mat{D} is a diagonal matrix  determined by the TES current, and $\mat{R}_0$ is representable as a stationary ARMA noise process.

%=============================================================================
\section{Conclusions}
\label{sec:conclusions}
%=============================================================================

Multi-pulse fitting is a technique to extend the virtues of optimal filtering into the regime of high photon rates. It fits simultaneously for the pulse heights of multiple piled-up pulses in an extended period longer than the usual optimal-filter length. Because pulses are permitted to overlap in time, the technique softens the usual requirement for long periods of isolation before and after any single valid pulse. 

Just as optimal filtering does, multi-pulse fitting takes advantage of inverse-noise weighting to achieve the maximum possible signal-to-noise ratio consistent with the data and the model assumptions. While this weighting is computationally expensive in the general case, we find success in replacing the full estimated noise-covariance function with a simplified approximation to it, a low-order ARMA model. 

%Accomplishments / measurements
The multi-pulse fitting technique has been applied to measurements from a single TES microcalorimeter operating at 6\,keV and to an array operating at 400 to 600\,eV. In the former case, the energy resolution remains better than 5\,eV at the Mn K$\alpha$ line for rates up to 100\,cps. At the lower energy, the resolution of an array of 2.5\,eV devices remains as low as 4.2\,eV even up to 262\,cps per sensor, and using a full 80\,\% of the pulses. The nitrogen fluorescence line complex of an ammonium nitrate target exhibits 2.5\,eV resolution at 46\,cps and 96\,\% pulse efficiency.

%Overview
We have shown results using a multi-pulse fitting technique on microcalorimeter measurements in the regime of very high photon-count rates and sensor linearity. Although nonlinearity and computational costs are not fully solved problems, we expect that this concept can be usefully adapted to a wide range of microcalorimeter data sets and believe it to be a promising step toward an analysis chain optimized for the demanding requirements of future x-ray space missions.

\acknowledgments
%{\bf Acknowledgments:}
We are grateful to our NIST electronics and microfabrication colleagues for preparation of outstanding microcalorimeter detectors, SQUID multiplexing electronics, and warm analog and digital readout systems. Colin Fitzgerald helped with the shipment of the spectrometer to the NSLS and its construction there. Jens Uhlig assisted with operations at the NSLS.
We had helpful conversations on pulse-analysis topics with Doug Bennett and Kent Irwin at NIST and Simon Bandler, Harvey Moseley, Dale Fixsen, and Sarah Busch of NASA's Goddard Space Flight Center. We thank the technical and computing staff of the National Synchrotron Light Source and of Brookhaven National Laboratory for their contributions.
This work was supported by the NIST Innovations in Measurement Science program, by NASA through the grant ``Demonstrating Enabling Technologies for the High-Resolution Imaging Spectrometer of the Next NASA X-ray Astronomy Mission'', NASA NNH11ZDA001N-SAT, and by an ARRA Senior Research Fellowship from NIST (JF).

\appendix
%=============================================================================
\section{Whitening ARMA(1,1) noise}
\label{appendix:arma11}

Section~\ref{sec:factoring_arma11} gives partial results for whitening data when Equation~\ref{eq:arma11} gives the noise covariance function. That is, it covers the case when the noise is described by an ARMA(1,1) process and its covariance is a decaying exponential plus a delta function at time 0.  The product $\mat{S}\equiv\mat{\Phi}\mat{R}\mat{\Phi}^T$ is a symmetric tridiagonal matrix with values
\begin{align}
S_{11} &= w+a \nonumber \\
S_{jj} &= w+a+\phi^2(w-a) \hspace{5mm} j\ge 2 \nonumber \\
S_{j\pm1,j} &= -w\phi. \nonumber 
\end{align} 

The Cholesky factorization $\mat{S}=\mat{B}\mat{B}^T$ is straightforward. Label the values on the diagonal of \mat{B}  $\{d_1,d_2,\dots d_n\}$ and on the subdiagonal  $\{e_1,e_2,\dots e_{n-1}\}$. They can be computed from the recursion:
\begin{align}
d_1 &= \sqrt{w+a} \nonumber \\
e_j &= -w\phi/d_j  \nonumber  \\
d_j &= \sqrt{w+a+\phi^2(w-a)-e^2_{j-1}}\nonumber 
\end{align}
and stored for later use. 

Equation~\ref{eq:arma11_whitener} shows that whitening the vector \vec{v} to yield $\vec{w}=\mat{W}\vec{v}$ requires solving $\mat{B}\vec{w} = \vec{y} = \mat{\Phi}\vec{v}$. A recursion for \vec{y} and \vec{w} is
\begin{align}
y_1 &= v_1 \nonumber \\
y_j &= v_j-\phi v_{j-1} \hspace{20mm} j\ge2 \nonumber \\
w_1 &= y_1/d_1 = v_1/d_1 \nonumber \\
w_j &= (y_j-e_{j-1}w_{j-1})/d_j  \hspace{10mm} j\ge2 \nonumber \\
&= (v_j - \phi v_{j-1} - e_{j-1}w_{j-1})/d_j .\nonumber 
\end{align}
This gives the whitened data \vec{w} in terms of the raw data \vec{v} and the elements of the Cholesky factor of \mat{S}.

The procedure for whitening (``inverting'') an ARMA($p,q$) process is similar, though necessarily more complicated than the $p=q=1$ case given in this Appendix.

%=============================================================================

\bibliography{multipulse_fitting}

\end{document}